\documentclass[final,3p,times]{elsarticle}
\usepackage{epstopdf}
\usepackage{graphicx}
\usepackage{algorithm}
\usepackage{algorithmic}
\usepackage{multirow}
\usepackage{amsmath}
\DeclareMathOperator*{\argmin}{argmin}
\DeclareMathOperator*{\argmax}{argmax}
\DeclareMathOperator*{\median}{median}
\DeclareMathOperator*{\mean}{mean}
\DeclareMathOperator*{\rank}{rank}
\usepackage{amssymb}
\usepackage{amsthm}

\journal{Computer Networks}

\begin{document}

\begin{frontmatter}
\title{Structural Analysis of Network Traffic Matrix via Relaxed Principal Component Pursuit\tnoteref{label1}}
\tnotetext[label1]{This work is supported by the National Natural Science Foundation of China (Grant No. 61073013)
and the Aviation Science Foundation of China (Grant No. 2010ZA04001)}
\author[1,2]{Zhe Wang\corref{cor1}}
\ead{wangzhe@cse.buaa.edu.cn}
\cortext[cor1]{Corresponding author.}
\author[1]{Kai Hu}
\ead{hukai@buaa.edu.cn}
\author[1,2]{Ke Xu}
\ead{kexu999@gmail.com}
\author[1,2]{Baolin Yin}
\ead{yin@nlsde.buaa.edu.cn}
\author[3]{Xiaowen Dong}
\ead{xiaowen.dong@epfl.ch}
\address[1]{School of Computer Science and Engineering, Beihang University, Beijing 100191, China}
\address[2]{State Key Laboratory of Software Development Environment, Beihang University, Beijing 100191, China}
\address[3]{Signal Processing Laboratories (LTS4 / LTS2), Ecole Polytechnique F\'{e}d\'{e}rale de Lausanne (EPFL), CH-1015 Lausanne, Switzerland}

\begin{abstract}
The network traffic matrix is widely used in network operation and management.
It is therefore of crucial importance to analyze the components and the structure of the network traffic matrix,
for which several mathematical approaches such as Principal Component Analysis (PCA) were proposed.
In this paper, we first argue that PCA performs poorly for analyzing traffic matrix that is polluted by large volume anomalies,
and then propose a new decomposition model for the network traffic matrix.
According to this model,
we carry out the structural analysis by decomposing the network traffic matrix into three sub-matrices, namely, the deterministic traffic, the anomaly traffic and the noise traffic matrix,
which is similar to the Robust Principal Component Analysis (RPCA) problem previously studied in \cite{Ma1}.
Based on the Relaxed Principal Component Pursuit (Relaxed PCP) method and the Accelerated Proximal Gradient (APG) algorithm,
we present an iterative approach for decomposing a traffic matrix,
and demonstrate its efficiency and flexibility by experimental results.
Finally, we further discuss several features of the deterministic and noise traffic.
Our study develops a novel method for the problem of structural analysis of the traffic matrix,
which is robust against pollution of large volume anomalies.
\end{abstract}

\end{frontmatter}

\section{Introduction}\label{Introduction}
\subsection{The Internet Traffic Data}\label{}
The Internet traffic data is considered as a significant input for network operation and management.
To monitor and analyze traffic data efficiently is one of the most important problems in the research field of network measurements.
In general, there are two levels of traffic data: the packet-level data and the flow-level data.
With the rapid growth of the Internet scale and link transmitting capability, on the one hand,
it is usually infeasible to collect and process the complete packet-level data;
On the other hand, the coarse flow-level data obtained by packet sampling often contains enough network information and has become increasingly popular in recent studies.
As an example, one type of widely used flow-level traffic data is the IP flows collected in each router by the Netflow protocol.
Roughly speaking, each IP flow is a sequence of packets sharing the same Source/Destination IP addresses,
Source/Destination port numbers and transport protocol during certain time intervals.
However, in large scale networks, the volume of IP flow data is still too huge for storage and processing.
For instance, the one-month IP flow data of the GEANT backbone network is about 150 GB \cite{geant},
which makes important applications such as anomaly detection impractical.
Therefore, it is necessary to further compress the IP flow data,
mainly by means of flow sampling and aggregation.

The network traffic matrix is computed by IP flow aggregation,
which records how much data is transmitted between each Original-Destination (OD) pair during different time intervals.
For each OD pair $(k_{1},k_{2})$,
the original point $k_{1}$ and the destination point $k_{2}$ are both the Points of Presence (PoP) in the network,
and we aggregate all the IP flows which enter the network at $k_{1}$ and exit at $k_{2}$ to represent the OD flow corresponding to $(k_{1},k_{2})$.
The research topics in traffic matrix analysis mainly include:
$(1)$ To estimate a traffic matrix accurately;
$(2)$ To generate synthetic traffic matrix;
$(3)$ To utilize a traffic matrix effectively for measurement applications,
such as anomaly detection and routing optimization.
These topics require a deep understanding of the components and structure of the traffic matrix.
In this paper, we carry out structural analysis of the traffic matrix by studying different traffic components that constitute the traffic matrix.

\subsection{Traffic Matrix and Its Structural Analysis}\label{}
Considering the network traffic with $p$ OD flows during $t$ time intervals,
the corresponding traffic matrix $X$ is a $t\times p$ non-negative matrix.
For each integer $1\leq j\leq p$,
the $j$-th column $X_{j}$ of $X$ is the traffic time series for the $j$-th OD flow;
for each integer $1\leq i\leq t$,
the $i$-th row $X(i,:)$ of $X$ is the traffic snapshot of all the OD flows during the $i$-th time interval.
According to the datasets adopted in this paper, we suppose $t>p$.
In real network measurements,
the Netflow protocol consumes much CPU resources;
some PoP routers might not support Netflow;
and the IP flow data might get lost during the transmission.
These limitations make the collection of a complete traffic matrix a challenging task,
therefore many estimation algorithms by means of indirect measurements (such as link traffic data collected by the SNMP protocol) were proposed in the literature.
In recent studies, the error of the third generation estimation algorithms have been decreased to below $10\%$.
In this paper, however, we do not concentrate on the estimation problem. Instead,
we perform our experiments based on real world datasets of traffic matrices.
These datasets are collected from the Abilene networks (in the U. S.) and the GEANT networks (in Europe),
which are available from \cite{data1} and \cite{data2}, respectively.

In general, a traffic matrix is a combination of different classes of network traffic.
In network operation and management,
people usually need information on all classes of traffic,
such as the deterministic traffic and the anomaly traffic.
In this paper, we study the structural analysis problem of a traffic matrix,
which is to accurately decompose the traffic matrix into sub-matrices that correspond to different classes of traffic,
hence explore in detail various features of the network traffic.

The most widely used approach for traffic matrix analysis is Principal Component Analysis (PCA) and its variants.
For example, Lakhina et al. \cite{Lakhina1} first introduced the PCA method in the studies of traffic matrices,
and they found that traffic matrices can be well approximated by a few principal components that correspond to the largest singular values of the matrices. Therefore, they claimed that traffic matrices usually have \emph{low effective dimensions}.
They further introduced the concept of \emph{eigenflow} and the eigenflow classification method,
discussed the distribution pattern of different eigenflow classes,
and proposed a method to decompose each OD flow time series according to the classification results.
These ideas were further developed in their later work \cite{Lakhina2},
in which they presented the PCA-subspace method for volume anomaly detection,
and decomposed the link traffic matrix into two sub-matrices that correspond to a normal subspace and an anomaly subspace, respectively.
During each time interval,
the norm of the traffic volume that corresponds to the anomaly subspace was compared with the Q-statistic threshold,
whose result was used to infer the existence of anomalies in the network.

After Lakhina's studies, many researchers have enriched the PCA-based methods for traffic matrix analysis.
Huang et al. \cite{Huang} proposed a distributed PCA method for volume anomaly detection,
which considered the trade-off between the detection accuracy and the data communication overhead.
Zhang et al. \cite{Zhang1} extended the classical PCA method
and argued that large volume anomalies should be extracted via both spatial and temporal approaches,
which were later named as \emph{Network Anomography}.
Based on the fact that traffic matrices often have low effective dimensions,
Soule et al. \cite{Soule1} proposed a new principal component method for traffic matrix estimation,
and the experiments demonstrated that their approach has a lower estimation error compared to most of the previous methods,
such as the tomogravity method and the route change method.

However, recent studies have shown some limitations of the PCA-based methods.
Ringberg et al. \cite{Ringberg} experimented the PCA-based anomaly detection method,
suggesting that its efficiency is very sensitive to the choice of parameters,
such as the number of principal components in the normal subspace,
the value of detection threshold,
and the level of traffic aggregation.
In addition, they found that large volume anomalies might pollute the normal subspace,
which could lead to high false positive rate in anomaly detection.
Ohsita et al. \cite{Ohsita} argued that
the traffic matrix estimated by network tomography is not a proper input for the PCA-based anomaly detectors.
Since most estimation methods are designed for an anomaly-free traffic matrix,
the estimation error might increase when the network traffic contains large volume anomalies.
Instead, they suggested estimating the increased traffic matrix and obtained a high attack-detection rate.
Their research also indicated the strict requirements of input traffic matrix for the PCA-based methods.
More recently, Rubinstein et al. \cite{Rubinstein} proved that attackers could significantly degrade the performance of the PCA-based anomaly detectors simply by adding chaff volumes before real attacks,
and designed an anomaly detector that is more robust against these attacks.

\subsection{Main Contributions of This Paper}\label{contributions}
As mentioned above,
although it has been extensively studied before,
PCA-based methods still have limitations for traffic matrix analysis and related applications.
One important drawback is that, when the traffic matrix is corrupted by large volume anomalies,
the resulting principal components will be significantly skewed from those in the anomaly-free case.
This prevents the subspace-based methods from accurately decomposing the total traffic into normal traffic and anomaly traffic,
and decreases the efficiency of PCA-based anomaly detectors.
However, to our knowledge there are only a few existing methods for analyzing a traffic matrix with large volume anomalies.
This is going to be the focus of this paper,
where we address the problem of structural analysis of polluted traffic matrices.
The main contribution of our paper is two-fold:

$(1)$ As the basic assumption behind the subspace-based methods is that each eigenflow can be exactly classified,
it is an interesting question whether those classification method still perform well for a polluted traffic matrix.
Specifically,
(i) is there an eigenflow that satisfies more than one classification criterion of the eigenflow classes?
(ii) is there an eigenflow that satisfies no classification criterion of the eigenflow classes?
(iii) does the distribution pattern of eigenflows maintain for a polluted traffic matrix?
We discuss these problems in Section \ref{Classical PCA}, where we use PCA for the structural analysis of real world traffic matrices,
which usually have large volume anomalies.

$(2)$ As the PCA-based structural analysis performs poorly when the traffic matrix contains large volume anomalies,
it is necessary to provide a new analysis tool that is suitable for polluted traffic matrices.
In Sections 3 and 4,
we propose a new decomposition model for the traffic matrix based on empirical network traffic data,
and formalize the mathematical definition of the structural analysis problem.
This motivates us to discover the equivalence between structural analysis and the Robust Principal Component Analysis (RPCA) problem.
We then design a decomposition algorithm based on the Relaxed Principal Component Pursuit (Relaxed PCP) method,
which is suitable for solving the RPCA problem.
Using this algorithm, we are able to obtain a proper traffic decomposition for the polluted traffic matrices in our experiments.
Finally, we analyze several properties of the sub-matrices from the decomposition of the traffic matrix in detail in Section 5.

\section{PCA for the Structural Analysis of Polluted Traffic Matrix}\label{Classical PCA}
\subsection{The Classical PCA Method}\label{}
PCA is widely used in high dimensional data analysis,
where the redundant high dimensional data can be approximated by a low dimensional representation.
In our study, we consider each row vector of the traffic matrix $X\in \mathbb{R}^{t\times p}$ as a data point in $\mathbb{R}^{p}$,
thus $X$ contains $t$ data points.
Following the common approach in \cite{Lakhina1}\cite{Lakhina2},
we normalize each OD flow vector (columns of $X$) to have zero mean before performing PCA:
\begin{equation}
X_{j}=X_{j}-\mean(X_{j}) \quad j=1, 2, ..., p.
\end{equation}

PCA can be viewed as a coordinate transformation process,
where the data points have been transformed from the original coordinate system to a new coordinate system.
All the unit vectors of the new coordinate system are represented as $\{v_{i}\}_{i=1}^{p}$,
and $v_{i}$ is called the $i$-th principal component vector.
The first principal component vector $v_{1}$ captures the maximum variance (energy) of the original traffic matrix $X$:
\begin{equation}
v_{1}=\argmax_{\parallel v \parallel =1}\parallel Xv \parallel .
\end{equation}

For each integer $k\geq 2$, suppose we have obtained the first $k-1$ principal component vectors,
the $k$-th principal component vector $v_{k}$ then captures the maximum variance of the residual traffic matrix,
which is the difference between the original traffic matrix $X$ and its mappings onto the first $k-1$ principal component vectors:
\begin{equation}
v_{k}=\argmax_{\parallel v \parallel =1}\parallel (X - \sum_{i=1}^{k-1}Xv_{i}v_{i}^{T})v\parallel .
\end{equation}

Following this progress,
all the principal component vectors are defined iteratively.
It is easy to show that $\{ v_{i}\}_{i=1}^{p}$ form an orthogonal basis of $\mathbb{R}^{p}$.
Thus the traffic matrix can be decomposed as:
\begin{eqnarray}\label{SVD1}
X & = & X[v_{1} \ v_{2} \ ... \ v_{p}][v_{1} \ v_{2} \ ... \ v_{p}]^{T}=\sum_{i=1}^{p}Xv_{i}v_{i}^{T} \nonumber \\
& =& \sum_{i=1}^{p}\parallel Xv_{i}\parallel \frac{Xv_{i}}{\parallel Xv_{i}\parallel} v_{i}^{T} = \sum_{i=1}^{p} \parallel Xv_{i}\parallel u_{i}v_{i}^{T},
\end{eqnarray}
where
\begin{equation}
u_{i}=\frac{Xv_{i}}{\parallel Xv_{i}\parallel} \quad i=1, 2, ..., p
\end{equation}
is a unit vector in $\mathbb{R}^{t}$,
which is called the $i$-th eigenflow corresponding to $v_{i}$ \cite{Lakhina1}.

Following basic matrix theory,
we can show that the principal component vectors $\{v_{i}\}_{i=1}^{p}$ are the eigenvectors of the matrix $X^{T}X$,
sorted by the corresponding eigenvalues $\{\lambda_{i}\}_{i=1}^{p}$ in a descending order:
\begin{equation}
X^{T}Xv_{i}=\lambda _{i}v_{i} \quad i=1, 2, ..., p,
\end{equation}
where $\lambda_{1}\geq \lambda_{2} \geq ... \geq \lambda_{p} \geq 0$.
Furthermore, $\sigma_{i}=\sqrt{\lambda_{i}}$ is called the $i$-th singular value of $X$.
Therefore, $\{ v_{i}\}_{i=1}^{p}$ can be found by computing the eigenvectors of $X^{T}X$.

Since $\parallel Xv_{i}\parallel = \sqrt{v_{i}^{T}X^{T}Xv_{i}} = \sqrt{\lambda_{i}v_{i}^{T}v_{i}}=\sigma_{i}$,
Equation (\ref{SVD1}) can be rewritten as:
\begin{equation}\label{SVD2}
X=\sum_{i=1}^{p}\sigma_{i}u_{i}v_{i}^{T}.
\end{equation}
Equation (\ref{SVD2}) is called the Singular Value Decomposition (SVD) of $X$.
By Eckart-Young theorem \cite{Stewart},
for each integer $1\leq r\leq p$,
$A_{r}=\sum_{i=1}^{r}\sigma_{i}u_{i}v_{i}^{T}$ is the best rank-$r$ approximation of $X$:
\begin{equation}\label{opti}
A_{r}=\argmin_{\rank(A)\leq r} \parallel X-A\parallel_{F},
\end{equation}
where $\|\cdot\|_{F}$ denotes the Frobenius matrix norm.
Each data point $X(s,:)^{T}\in \mathbb{R}^{p} \ (1\leq s\leq t)$
can then be approximated as
\begin{equation}
X(s,:)^{T}\approx \sum_{i=1}^{r}\sigma_{i}u_{i}(s)v_{i},
\end{equation}
where $\{ \sigma_{i}u_{i}(s)\}_{i=1}^{r}$ are the first $r$ coefficients of $X(s,:)^{T}$ under the new coordinate system.
Therefore, PCA can be viewed as a technique for dimensionality reduction.

\subsection{Eigenflows of The Polluted Traffic Matrix}\label{classification}
In the following experiments,
we adopt two widely used traffic matrix datasets, one from the Abilene network and the other from the GEANT network.
Abilene is a Internet2 backbone network with 12 PoPs (144 OD flows),
and GEANT is a pan-European research network with 23 PoPs (529 OD flows).
In this paper,
the minimal time interval in the Abilene dataset is 5 minutes;
the minimal time interval in the GEANT dataset is 15 minute, since the flow files are written in 15 minutes.
In fact, there exists flow data with finer time scale.
For example, the minimal time interval of another GEANT traffic dataset is precise to one second in \cite{Diot}.
The Abilene dataset contains 24 weeks' traffic records
from March 1, 2004 to September 10, 2004,
and each week's data is represented by a traffic matrix.
Here we select the traffic matrices that correspond to the first 8 weeks (denoted as $\textrm{X}01\sim \textrm{X}08$).
For the GEANT dataset,
traffic records are not complete for certain days.
Since most of the researchers study the weekly traffic matrix,
we choose a subset of the GEANT dataset:
the four consecutive weeks' traffic matrices from March 28, 2005 to April 24, 2005
(denoted as $\textrm{Y}01\sim \textrm{Y}04$).
Each traffic matrix in the Abilene dataset consists of 2016 rows (time steps),
while each GEANT traffic matrix consists of 672 rows.

Notice that for each of these traffic matrices,
there are a small number of OD flow time series that contain a large percentage of zero entries.
This usually means that these OD flows are not stable,
therefore we delete them in the experiments.
For each traffic matrix in the Abilene dataset,
we delete the 23 OD flows whose source or destination PoP is "ATLA-M5"
(thus the number of OD flows actually used is 121);
For each traffic matrix in the GEANT dataset,
we delete those OD flows that have more than 50\% zero entries
(the number of OD flows actually used is between 457 and 483).
Table 1 summarizes the datasets used in our experiments,
and Table 2 (the first column) shows the number of OD flows actually used for each traffic matrix.

\begin{table}[!htbp]
\begin{tabular*}{\columnwidth}{@{}llllll}
\multicolumn{5}{l}{\footnotesize \textbf{Table 1}}\\
\multicolumn{5}{l}{\footnotesize Datasets used in the experiments}\\[5pt]
\hline
\footnotesize Name &\footnotesize $\#$OD Flows/Actually Used &\footnotesize Time Interval &\footnotesize Time Steps &\footnotesize Peroid\\
\hline
\footnotesize Abilene &\footnotesize 144/121 &\footnotesize 5 minutes &\footnotesize 2016 &\footnotesize 8 weeks\\
\footnotesize GEANT &\footnotesize 529/457-483 &\footnotesize 15 minutes &\footnotesize 672 &\footnotesize 4 weeks\\
\hline
\end{tabular*}
\end{table}

In Lakhina's original study \cite{Lakhina1},
all the eigenflows of a traffic matrix can be classified into three types:
First, the eigenflows that exhibit distinct periodical patterns are called \emph{d-eigenflow} (for "deterministic"),
since they reflect the diurnal activity in the network traffic,
as well as the difference between weekday and weekend activities;
Second, the eigenflows that represent strong, short-lived spikes are called \emph{s-eigenflow} (for "spike"),
as they capture the occasional traffic bursts and dips which are usually reported;
Third, the eigenflows that roughly have a stationary and Gaussian behavior are called \emph{n-eigenflow} (for "noise"),
since they capture the remaining random variation that arises due to multiplexing of many individual traffic sources.

In this paper, we follow the classification criteria in \cite{Lakhina1} for both d-eigenflow and s-eigenflow.
The original classification criterion for n-eigenflow in \cite{Lakhina1} is to compare the qq-plot of eigenflow's distribution with the normal distribution; However, it is not considered as a quantitative method. Therefore, we use another classification criterion from the Kolmogorov-Smirnov (K-S) test instead. Suppose $u_{j}$ is an eigenflow of the traffic matrix $X$, we classify it according to the following three criteria:

  (1) d-eigenflow:
  Let $H$ denote the set of period parameters measured in hours.
  For each element $h\in H$,
  we compute the Fourier power spectrum $\widetilde{u_{j}}(h)$ of $u_{j}$:
  \begin{equation}\label{DFT}
  \begin{cases}
  \widetilde{u_{j}}(h) & = \Big| \sum_{k=0}^{t-1}u_{j}(k+1)\cdot \exp(-\omega ki)\large\Big|^{2}\Big/t \nonumber \\
  \omega & = 2\pi/T \\
  T & = 60h/t_{0}
  \end{cases},
  \end{equation}
  where $T$ is the period of Fourier transform
  and $t_{0}$ is the length of time interval (measured in minutes).
  In this paper, $H=\{ k\}_{k=1}^{10}\bigcup \{ 2k\}_{k=6}^{25}$.
  If $\{12,24\}\bigcap \argmax_{k\in H}\{\widetilde{u_{j}}(k)\}\neq \phi$,
  $u_{j}$ satisfies the criterion of d-eigenflow and we classify it as d-eigenflow;

  (2) s-eigenflow: Let $\sigma$ denote the standard deviation of $u_{j}$.
  If $u_{j}$ has at least one entry outside the interval $[\mean(u_{j})-5\sigma,\mean(u_{j})+5\sigma]$,
  $u_{j}$ satisfies the criterion of s-eigenflow and we classify it as s-eigenflow;

  (3) n-eigenflow: We use the K-S test to verify the normal distribution of $u_{j}$.
  If the null hypothesis (Normal Distribution) is not rejected at $5\%$ significance level,
  $u_{j}$ satisfies the criterion of n-eigenflow and we classify it as n-eigenflow.

In order to evaluate the completeness (each eigenflow has to be classified into at least one class) and orthogonality (the same eigenflow must not be classified into more than one class at the same time) of eigenflow classification,
we further define the following two concepts:
\begin{itemize}
  \item indeterminate eigenflow: eigenflows that satisfy more than one classification criterion;
  \item non-determinate eigenflow: eigenflows that satisfy no classification criterion.
\end{itemize}

\begin{table*}[!bhtp]
\begin{tabular*}{\textwidth}{@{}llllllll}
\multicolumn{8}{l}{\footnotesize \textbf{Table 2}}\\
\multicolumn{8}{l}{\footnotesize Eigenflow classification using PCA}\\[5pt]
\hline
\footnotesize Traffic matrix name\quad &\footnotesize $\#$ Satisfy &\footnotesize $\#$ Satisfy &\footnotesize $\#$ Satisfy &\footnotesize $\#$ non-determinate &\footnotesize $\#$ Indeterminate &\footnotesize $\#$ Classified &\footnotesize Unclassified\\
\footnotesize ($\#$ OD flows used) &\footnotesize d-eigenflow &\footnotesize s-eigenflow &\footnotesize n-eigenflow &\footnotesize eigenflow &\footnotesize eigenflow &\footnotesize eigenflow &\footnotesize energy rate \\
\hline
\footnotesize X01 (121) &\footnotesize 10 &\footnotesize 42 &\footnotesize 87 &\footnotesize 10 &\footnotesize 26 &\footnotesize 85  &\footnotesize 83.08\% \\
\footnotesize X02 (121) &\footnotesize 9  &\footnotesize 46 &\footnotesize 80 &\footnotesize 10 &\footnotesize 22 &\footnotesize 89  &\footnotesize 82.59\% \\
\footnotesize X03 (121) &\footnotesize 14 &\footnotesize 78 &\footnotesize 47 &\footnotesize 12 &\footnotesize 29 &\footnotesize 80  &\footnotesize 11.08\% \\
\footnotesize X04 (121) &\footnotesize 10 &\footnotesize 62 &\footnotesize 74 &\footnotesize 5  &\footnotesize 28 &\footnotesize 88  &\footnotesize 94.67\% \\
\footnotesize X05 (121) &\footnotesize 8  &\footnotesize 73 &\footnotesize 62 &\footnotesize 9  &\footnotesize 30 &\footnotesize 82  &\footnotesize 46.09\% \\
\footnotesize X06 (121) &\footnotesize 15 &\footnotesize 52 &\footnotesize 83 &\footnotesize 9  &\footnotesize 33 &\footnotesize 79  &\footnotesize 0.32\%  \\
\footnotesize X07 (121) &\footnotesize 6  &\footnotesize 63 &\footnotesize 75 &\footnotesize 9  &\footnotesize 30 &\footnotesize 82  &\footnotesize 9.39\%  \\
\footnotesize X08 (121) &\footnotesize 10 &\footnotesize 53 &\footnotesize 71 &\footnotesize 11 &\footnotesize 24 &\footnotesize 86  &\footnotesize 98.49\% \\
\footnotesize           &\footnotesize    &\footnotesize    &\footnotesize    &\footnotesize    &\footnotesize    &\footnotesize     &\footnotesize         \\
\footnotesize Y01 (483) &\footnotesize 17 &\footnotesize 54 &\footnotesize 469 &\footnotesize 5 &\footnotesize 59 &\footnotesize 419 &\footnotesize 76.67\% \\
\footnotesize Y02 (465) &\footnotesize 5 &\footnotesize 66 &\footnotesize 453 &\footnotesize 6 &\footnotesize 65 &\footnotesize 394 &\footnotesize 77.70\% \\
\footnotesize Y03 (465) &\footnotesize 7 &\footnotesize 47 &\footnotesize 454 &\footnotesize 5 &\footnotesize 48 &\footnotesize 412 &\footnotesize 43.19\% \\
\footnotesize Y04 (457) &\footnotesize 12 &\footnotesize 62 &\footnotesize 444 &\footnotesize 5 &\footnotesize 66 &\footnotesize 386 &\footnotesize 82.42\% \\
\hline
\end{tabular*}
\end{table*}

Next, we apply PCA to compute the principal component vectors and eigenflows of each traffic matrix.
Following the classification criteria proposed above,
we summarize the classification results of eigenflows in Table 2.
Here we define the \emph{unclassified energy rate} as the percentage of energy captured by the principal component vectors
that correspond to either indeterminate or non-determinate eigenflows.
Since the energy captured by one principal component vector is proportional to the square of the corresponding singular value, we have:
\begin{equation}
\text{unclassified energy rate} = \frac{\sum_{k\in UEID}\lambda_{k}}{\sum_{i=1}^{p}\lambda_{i}},
\end{equation}
where the union of unclassified eigenflow ID (\emph{UEID}) is
\begin{equation}
UEID=\{\ k \ | \ u_{k} \ \text{is indeterminate or non-determinate} \}.
\end{equation}

Based on the classification results of the twelve weekly traffic matrices described above,
we make the following observations on traffic matrices that are possibly polluted by large volume anomalies:

$(1)$ One the one hand, only a small number of eigenflows satisfy the classification criterion of d-eigenflow
(usually less than 20),
and most of them correspond to large singular values;
On the other hand, there is a considerable number of eigenflows satisfying the classification criterion of s-eigenflow and n-eigenflows.
The proportion of each eigenflow class varies from one traffic matrix to another.
These are similar to Lakhina's experimental results presented in \cite{Lakhina1}.

$(2)$ The PCA-based eigenflow classification method shows serious limitations in terms of classification completeness and orthogonality.
Specifically, a large number of eigenflows can not be exactly classified into one eigenflow class.
Furthermore, some unclassified eigenflows correspond to large singular values,
and the unclassified energy rate is larger than 70\% for seven of the twelve traffic matrices.
These results are not consistent with those in Lakhina's study \cite{Lakhina1},
in which the non-determinate eigenflows do not exist and the indeterminate eigenflows only contribute little energy.
Although the authors in \cite{Lakhina1} argued that their classification method could be enhanced by heuristic mechanisms,
our experiments show that some unclassified eigenflows are essentially different from all the eigenflow classes,
hence the classification results can not be clearly improved
only by changing parameters or adopting heuristic algorithms.

$(3)$ For each traffic matrix in the Abilene dataset $(\textrm{X}01\sim \textrm{X}08)$,
the first six eigenflows (eigenflows corresponding to the six largest singular values) often contain some instances satisfying the classification criterion of s-eigenflow.
This does not happen in Lakhina's study \cite{Lakhina1},
where the first six eigenflows are exactly classified as d-eigenflows.
For the GEANT traffic matrices, the first six eigenflows do not satisfy the criterion of s-eigenflow in general, which can be explained by the fact that the anomaly volumes in the GEANT networks are not as large as that in the Abilene networks.

\begin{figure}[!htbp]
\centering
\scalebox{0.8}[0.8]{\includegraphics*{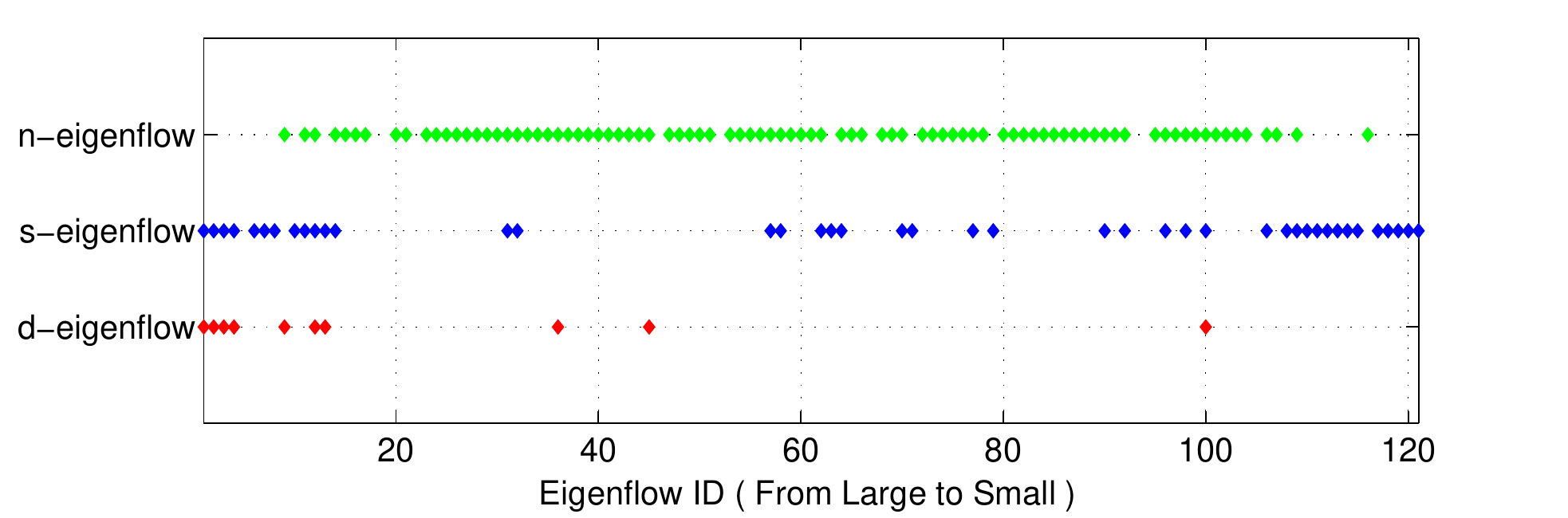}}
\begin{center}
\caption{Eigenflow classification result for the traffic matrix X01}\label{X01_cla}
\end{center}
\medskip
\end{figure}

\begin{figure}[!htbp]
\centering
\scalebox{0.85}[0.85]{\includegraphics*{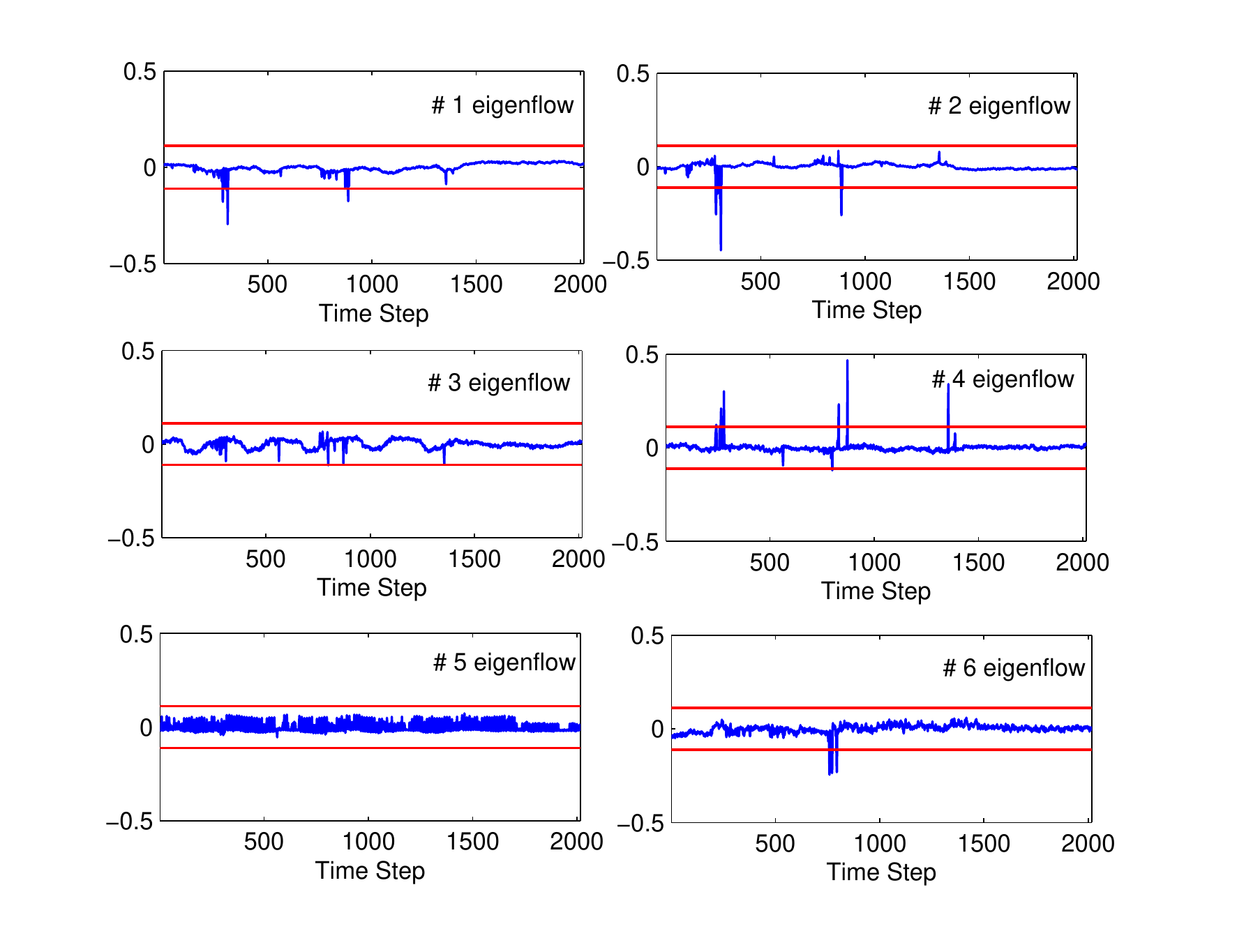}}
\begin{center}
\caption{The first six eigenflows of the traffic matrix X01}\label{X01_eig}
\end{center}
\medskip
\end{figure}

\begin{figure}[!htbp]
\centering
\scalebox{0.85}[0.85]{\includegraphics*{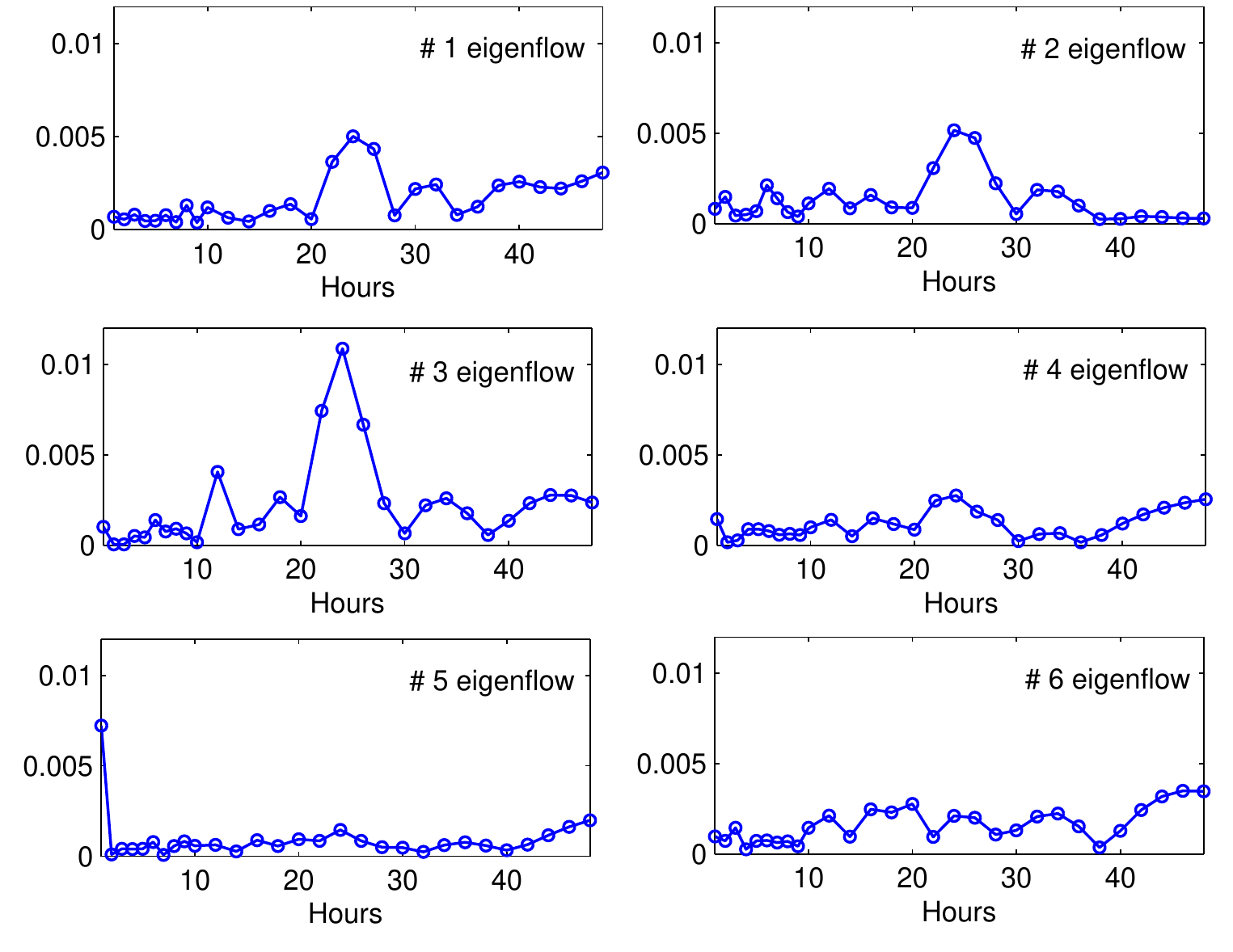}}
\begin{center}
\caption{Fourier power spectra of the first six eigenflows of X01}\label{X01_eig_spe}
\end{center}
\medskip
\end{figure}

Here we present the experimental result for the traffic matrix $\textrm{X}01$ as a case study.
Figure \ref{X01_cla} shows the classification result of the 121 eigenflows
(sorted from large to small by the corresponding singular values),
where each indeterminate eigenflow appears simultaneously in more than one classes whose classification criteria it satisfies.
The first six eigenflows and their Fourier power spectra are shown in blue in Figure \ref{X01_eig} and Figure \ref{X01_eig_spe}, respectively.
Each pair of red lines in Figure \ref{X01_eig} is the $+5\sigma/-5\sigma$ boundary for the corresponding eigenflow,
which is used for the inference of s-eigenflow.
It is clear that five out of six eigenflows in Figure \ref{X01_eig} satisfy the classification criterion of s-eigenflow.
However, for the first four eigenflows,
their Fourier power spectra all achieve the maximum values when the period parameter is equal to 24 hours,
suggesting that they satisfy the criterion of d-eigenflow.
Therefore,
we can view each of the first four eigenflows as a hybrid time series
mixed with the deterministic diurnal pattern and the short-lived anomaly pattern,
which is quite different from the three pre-defined eigenflow classes.
Changing parameters or using heuristic algorithms in classification would not help much in this case.
If we classify these four eigenflows as d-eigenflow,
the energy of anomaly traffic will be significantly underestimated,
which might increase the false negative rate of anomaly detection algorithms;
On the contrary, classifying them as s-eigenflow will lead to underestimation of normal network traffic,
which could prevent us from correctly decomposing the deterministic traffic component from a polluted traffic matrix.

From the discussions above, we can see that PCA-based method has limitations in eigenflow classification:
(i) for both the Abilene dataset and the GEANT dataset,
a large proportion of eigenflows could not be exactly classified into one eigenflow class;
(ii) for the Abilene dataset,
some eigenflows corresponding to the six largest singular values satisfy the classification criterion of s-eigenflow,
which would be problematic for us to isolate the deterministic traffic trend or to detect the anomaly traffic events.
This motivates us to propose a new model for the structural analysis of traffic matrix in the next section.

\section{Relaxed Principal Component Pursuit for the Structural Analysis of Polluted Traffic Matrix}\label{PCP_section}
\subsection{The Decomposition Model of Traffic Matrix}\label{Composition Model}
In this section we propose a new decomposition model for the traffic matrix data,
and discuss the mathematical nature of the structural analysis problem. Next,
based on the optimization process corresponding to PCA,
we explain intuitively the limitations of PCA-based method for analyzing a polluted traffic matrix.
This helps interpret the experimental results presented in Section \ref{classification} more deeply.

According to the empirical measurement data,
we suppose that there exist three classes of network traffic:
the deterministic traffic which shows diurnal pattern;
the anomaly traffic which appears rarely but involves large peak-like or block-like volumes;
the noise traffic that has small magnitude but appears in every OD flow during all the time intervals.
Formally speaking, we propose to decompose each traffic matrix into three sub-matrices:
\begin{equation}\label{A+E+N}
X=A+E+N,
\end{equation}
where $A$, $E$ and $N$ represent the deterministic traffic matrix,
the anomaly traffic matrix and the noise traffic matrix, respectively.
Each class of traffic has its own features,
based on which we make the following hypotheses:

$(1)$ The deterministic traffic is mainly contributed by periodical traffic.
This means that the periodical traffic time series of different OD flows have similar periods and phases,
and they mainly differ in magnitudes.
This implies that $A$ should be a low-rank matrix;

$(2)$ The anomaly traffic does not show up frequently,
which implies that $E$ should be a sparse matrix. However, its nonzero entries may have large magnitudes;

$(3)$ We assume that $N$ is a random matrix consists of independent zero-mean Gaussian random variables.
Each column of $N$ (the noise traffic time series of an OD flows) can be viewed as Gaussian random variables
which have the same variance.
Since the noise variance is usually proportional to the scale of the corresponding OD flow,
and different OD flows may have very different scales,
the Gaussian random variables corresponding to different columns may have different variances.

Equation (\ref{A+E+N}) and hypotheses $(1)-(3)$ constitute our decomposition model for the traffic matrix.
In practical measurements, only the traffic matrix $X$ can be observed,
thus the mathematical nature of the structural analysis problem is
to exactly decompose $X$ into three components, namely $A$, $E$ and $N$ in our decomposition model.

\medskip

Consider a simple case where we neglect the noise traffic matrix $N$ in (\ref{A+E+N}).
Suppose that $rank(A)=r_{0}$. According to Section 2.1,
$A_{r_{0}}=\sum_{i=1}^{r_{0}}\sigma_{i}u_{i}v_{i}^{T}$ is the best rank-$r_{0}$ approximation of $X$.
Then, one natural question is to ask whether $A=A_{r_{0}}$.
If this is true, the PCA method would achieve a decomposition of $X$ that is consistent with our decomposition model
($A_{r_{0}}=A$, hence $X-A_{r_{0}}=E$).
However, it has not been shown that $A$ is the solution to the optimization problem (\ref{opti}).
Considering the magnitude of nonzero entries in $E$,
our discussion is further divided into the following two cases:

$(1)$ If the anomaly traffic only has entries with small magnitude,
the Frobenius matrix norm $\|E\|_{F}=\|X-A\|_{F}$ would be small,
thus the deterministic traffic matrix $A$ is most likely to be the solution to problem (\ref{opti}).
In this case $A=A_{r_{0}}$, and these two matrices have the same eigenflows.
As the deterministic traffic is mainly contributed by diurnal traffic,
the first $r_{0}$ eigenflows of $A$ are usually d-eigenflows (since $rank(A)=r_{0}$).
Following the definition of $A_{r_{0}}$,
these eigenflows are also the first $r_{0}$ eigenflows of $X$.
Therefore, when the anomaly traffic has entries with small magnitude,
the PCA method performs well in eigenflow classification.

$(2)$ If the anomaly traffic has entries with large magnitude,
even though $E$ is a sparse matrix,
its Frobenius matrix norm $\|E\|_{F}=\|X-A\|_{F}$ would be large.
As a result, $A$ is usually not the solution to problem (\ref{opti}),
and $A_{r_{0}}$ contains a large amount of anomaly traffic.
Therefore, some of the first $r_{0}$ eigenflows of $X$ may satisfy the classification criterion of s-eigenflow.

The discussions above explain intuitively the experimental results presented in Section \ref{classification}.
The PCA-based eigenflow classification can be considered as a special method for traffic matrix decomposition.
However, when the traffic matrix is polluted by large volume anomalies,
PCA can not achieve a complete and orthogonal eigenflow classification,
and the PCA-based traffic matrix decomposition is inconsistent with the proposed decomposition model.

\subsection{Robust Principal Component Analysis and Principal Component Pursuit}\label{PCP_method}
Following the decomposition model proposed in Section \ref{Composition Model},
the structural analysis problem of the traffic matrix is to accurately decompose the original traffic matrix $X$
into a deterministic traffic matrix $A$, an anomaly traffic matrix $E$, and a noise traffic matrix $N$.
This is similar to the Low-rank Matrix Recovery problem,
which is also known as \emph{Robust Principal Component Analysis (RPCA)}\cite{Ma1}.

Recently, developments in the theory of \emph{Compressive Sensing} \cite{Donoho1}\cite{Candes1}
have attracted wide attentions in the field of information science.
Compressive Sensing theory states that:
If the signal has a sparse representation under some orthonormal bases or dictionaries,
it can then be recovered by far fewer samples or measurements than that are needed using traditional methods.
Partially motivated by this claim,
Ma et al. proposed the \emph{Principal Component Pursuit (PCP)} method for the RPCA problem.
They studied the approximate algorithms for PCP,
and applied it to different real world applications such as video background modeling \cite{Ma1},
face alignment \cite{Ma3},
and web document corpus analysis \cite{Ma4}.
Following the definitions in \cite{Ma1}\cite{Ma2}\cite{Ma5},
we briefly describe the RPCA problem and the PCP method as follows.
We assume that $X, A, E, N$ are real matrices in $\mathbb{R}^{t\times p}$;
$\Lambda(\cdot)$ denotes the support set of a matrix,
which is the union of non-zero positions of the matrix.

\noindent\textbf{Problem 1} (standard RPCA problem) Suppose that $X=A+E$,
where $A$ and $E$ are two unknown matrices.
Assume that $A$ is a low-rank matrix and $E$ is a sparse matrix,
the standard RPCA problem is to recover $A$ and $E$ from $X$.

The authors in \cite{Ma5} suggested that the standard RPCA problem can be formulated as the following optimization problem:
\begin{equation}\label{l0_opt}
\min_{A, \ E} rank(A)+\gamma\|E\|_{0} \quad s.t. \ A+E=X,
\end{equation}
where $\|\cdot\|_{0}=|\Lambda(\cdot)|$ is the degree of the support set, which is also called the $l_{0}$-norm of the matrix;
$\gamma$ is a positive parameter that balance the two competing terms.
Problem (\ref{l0_opt}) consists of two sub-problems, namely the low-rank matrix completion problem and the $l_{0}$-norm minimization problem.
Unfortunately, both of them are NP-hard,
which makes problem (\ref{l0_opt}) intractable in polynomial time.

On the one hand, thanks to the developments in Compressive Sensing theory,
a series of previous works suggested the equivalence between the $l_{0}$-norm minimization and the $l_{1}$-norm minimization problems;
On the other hand, recent research on the matrix completion problem \cite{Candes2}
studied the matrix nuclear norm $\|\cdot\|_{\ast}$
(for a matrix $A\in\mathbb{R}^{t\times p}$,
the nuclear norm $\|A\|_{\ast}=\sum_{k=1}^{p}\sigma_{k}(A)$ is defined as
the sum of its singular values $\{\sigma_{k}(A)\}_{k=1}^{p}$),
and indicated that the two optimization problems, namely,
the matrix rank minimization problem and
the problem of minimizing the matrix nuclear norm, usually produce similar results.
More importantly, both the $l_{1}$-norm minimization and the nuclear norm minimization problems are convex optimization problems hence can be solved efficiently.

The PCP method is used for solving a hybrid optimization problem that consists of $l_{1}$-norm minimization and nuclear norm minimization:
To relax the objective function in (\ref{l0_opt}),
we replace the $l_{0}$-norm with the $l_{1}$-norm,
and the rank with the matrix nuclear norm $\|\cdot\|_{\ast}$, respectively.
Candes et al. \cite{Ma1} proved that,
under surprisingly broad conditions,
"almost all" matrices of the form $X=A+E$, namely,
matrices that are the sums of a low-rank matrix $A$ and a sparse matrix $E$,
can be exactly decomposed into these two components by solving the following convex optimization problem:
\begin{equation}\label{PCP1}
\min_{A, \ E} \|A\|_{\ast}+\lambda\|E\|_{1} \quad s.t. \ A+E=X,
\end{equation}
where $\lambda>0$ is a regularization parameter.
In addition, they showed that $\lambda=1/\sqrt{\max(t,p)}$ is a proper choice for the parameter that is independent from $A$ and $E$,
and we follow this choice throughout the rest of the paper.

The standard RPCA problem assumes that $X$ is strictly equal to the sum of a low-rank matrix and a sparse matrix.
However, in many real world applications,
observational data often contains certain level of noise,
and it usually pollutes almost all the entries of the matrix.
One of the most common perturbations is the Gaussian white noise,
which leads to the generalized RPCA problem with Gaussian noise:

\noindent\textbf{Problem 2} (generalized RPCA problem with Gaussian noise) Suppose that $X=A+E+N$,
where $A$, $E$ and $N$ are unknown matrices.
Assume that $A$ is a low-rank matrix, $E$ is a sparse matrix, and $N$ is a random matrix
whose entries follow i.i.d. zero-mean Gaussian distributions
with $\|N\|_{F}<\delta$ for some positive $\delta$.
The generalized RPCA problem is to recover $A$ and $E$ from $X$ under the perturbation of $N$.

For Problem 2, Zhou et al. \cite{Ma2} generalized the PCP method to propose the Relaxed PCP method,
which is to solve the following optimization problem:
\begin{equation}\label{PCP2}
\min_{A, E} \|A\|_{\ast}+\lambda\|E\|_{1} \quad s.t. \ \|X-A-E\|_{F}\leq\delta .
\end{equation}
They proved that,
under the same conditions as that PCP requires,
for any realization of the Gaussian noise satisfying $\|N\|_{F}<\delta$,
the solution to the generalized RPCA problem (\ref{PCP2}) gives a stable estimation of $A$ and $E$ with high probability.

The assumptions in the generalized RPCA problem are similar to
the hypotheses on the traffic matrix in our decomposition model (\ref{A+E+N}).
In fact, we can multiply the columns of the traffic matrix by some constants
to make sure that the Gaussian random variables in the noise traffic matrix have the same variance.
Meanwhile, this multiplication preserves the rank of the deterministic traffic matrix and the sparsity of the anomaly traffic matrix.
Therefore, the Relaxed PCP method can be used for solving our structural analysis problem.

\subsection{The Accelerated Proximal Gradient Algorithm}\label{APG section}
The Relaxed PCP method used for solving the constrained optimization problem (\ref{PCP2})
is usually computationally expensive.
A more efficient way is to solve an equivalent unconstrained optimization problem instead,
using algorithms such as
Iterative Thresholding (IT) \cite{Ma1},
Augmented Lagrange Multiplier (ALM) and Accelerated Proximal Gradient (APG) \cite{Ma2}.
In this paper, we adopt the APG algorithm,
which solves the following unconstrained minimization problem:
\begin{equation}\label{APG}
\min_{A,E} \mu\|A\|_{\ast}+\mu\lambda\|E\|_{1}+\frac{1}{2}\|X-A-E\|_{F}^2,
\end{equation}
where $\frac{1}{2}\|X-A-E\|_{F}^2$ is the penalty function,
and $\mu>0$ is a Laugragian parameter.
It has been proved in \cite{Ma2} that with some proper choices of $\mu=\mu(\delta)$,
the solution to (\ref{APG}) is equivalent to the solution to (\ref{PCP2}).

As mentioned above,
the choice of the regularization parameter $\lambda$ follows that in \cite{Ma1} and \cite{Ma2}:
\begin{equation}\label{lambda}
\lambda=\frac{1}{\sqrt{\max(t,p)}}.
\end{equation}

\noindent For the Lagrangian parameter $\mu$,
it is chosen as $\sqrt{2\max(t,p)}\sigma$ and $(\sqrt{t}+\sqrt{p})\sigma$ in \cite{Ma2} and \cite{Candes3}, respectively,
where $\sigma$ is the variance of Gaussian noise matrix $N$.
These choices are motivated by neglecting the effect of the sparse matrix $E$:
if we set $E=0$ in problem (\ref{APG}),
the APG algorithm which solves this problem boils down to the \emph{Singular Value Thresholding} algorithm with total sampling \cite{Ma2}\cite{Candes4}.
In our case, since the anomaly traffic matrix might contribute a large proportion of energy,
these choices are not suitable.
Therefore we present a new choice of $\mu$ in this paper.
Considering the case when $A=0$,
problem (\ref{APG}) boils down to:
\begin{equation}\label{BP}
\min_{A,E}\mu\lambda\|E\|_{1}+\frac{1}{2}\|X-E\|_{F}^{2}.
\end{equation}
If we consider $X$ and $E$ as two column vectors of dimension $t\times p$,
problem (\ref{BP}) becomes the \emph{Basis Pursuit Denoising} problem first introduced in \cite{Donoho2}\cite{Donoho3}.
As $E$ is a sparse vector,
we follow \cite{Donoho2} to choose $\mu\lambda=\sigma\sqrt{2\log(tp)}$.
Since $\lambda$ is chosen as in (\ref{lambda}),
we compute $\mu$ accordingly as:
\begin{equation}\label{mu}
\mu=\sigma\sqrt{2\log(tp)\max(t,p)}.
\end{equation}

For each OD flow time series $X_{j}\in \mathbb{R}^{t}$ ($1\leq j\leq p$),
we need to estimate the variance $\sigma_{j}$ of the Gaussian noise traffic.
This is a well-studied signal processing problem.
We adopt the estimation method proposed in \cite{Donoho4}:
Given an orthonormal wavelet basis,
and let $W_{j}=\{a_{k}^{j}\}_{k=1}^{t/2}$ denote $X_{j}$'s wavelet coefficients at the finest scale,
$\sigma_{j}$ is estimated as the median absolute deviation of $W_{j}$
divided by 0.6745:
\begin{equation}\label{sigma_j}
\sigma_{j}=\frac{1}{0.6745}\median\{|a_{k}^{j}-\median(W_{j})|\},
\end{equation}
where $\median(\cdot)$ denotes the median value of a vector.
This estimation method is motivated by the empirical fact that, wavelet coefficients at the finest scale are,
with few exceptions, essentially pure noise.
In this paper, we adopt the Daubechies-5 wavelet basis.

Now we are ready to present the proposed Algorithm \ref{algo1} (see Appendix) for traffic matrix decomposition,
which is partially based on the noisy-free version of the APG algorithm in \cite{Ma6}.

\section{Experiments}\label{Experiments}
We decompose the twelve traffic matrices described in Section \ref{Classical PCA} using Algorithm \ref{algo1}
($\textrm{X}01-\textrm{X}08$ are from the Abilene dataset
and $\textrm{Y}01-\textrm{Y}04$ are from the GEANT dataset).
The detailed experimental results are summarized in Table 3,
where each row corresponds to the decomposition result for one traffic matrix.
From left to right, the columns of Table 3 represent:

$(1)$ Name of the Traffic matrix;

$(2)$ The rank of the deterministic traffic matrix;

$(3)$ The rank of the original traffic matrix;

$(4)$ The ratio of $(2)$ to $(3)$,
to evaluate the relative low-rank degree of the deterministic traffic matrix;

$(5)$ The $l_{0}$-norm of the anomaly traffic matrix;

$(6)$ $t\times p$, where $t$ and $p$ are the number of rows and columns of the traffic matrix, respectively;

$(7)$ The ratio of $(5)$ to $(6)$,
to evaluate the relative sparsity level of the anomaly traffic matrix;

$(8)$ The ratio of Frobenius norm of the noise traffic matrix to that of the corresponding original traffic matrix,
to evaluate how much energy is contained in the noise traffic matrix;

$(9)$ Number of iterations implemented in the APG algorithm for each traffic matrix.
In this paper, the tolerance parameter for the stopping criterion is set to $10^{-6}$;

$(10)$ Computational time of the implementation of the APG algorithm for each traffic matrix (in seconds).
In all the experiments, we use a commercial PC with 2.0GHz Intel Core2 CPU and 2.0GB RAM.

\begin{table*}[!htbp]
\begin{tabular*}{\textwidth}{@{}llllllllll}
\multicolumn{10}{l}{\footnotesize \textbf{Table 3}}\\
\multicolumn{10}{l}{\footnotesize Traffic matrix decomposition results using Relaxed PCP}\\[5pt]
\hline
\footnotesize traffic matrix
&\footnotesize rank$(A)$ &\footnotesize rank$(X)$ &\footnotesize rank$(A)$/rank$(X)$
&\footnotesize $\|E\|_{0}$ &\footnotesize $t\times p$ &\footnotesize $\|E\|_{0}/(t\times p)$
&\footnotesize $\|N\|_{F}/\|X\|_{F}$ &\footnotesize $\#$ iteration &\footnotesize computation time(s)\\
\hline
\footnotesize X01
&\footnotesize 10 &\footnotesize 121 &\footnotesize 0.0826
&\footnotesize 32766 &\footnotesize 243936 &\footnotesize 0.1343
&\footnotesize 0.1718  &\footnotesize 86  &\footnotesize 24\\
\footnotesize X02
&\footnotesize 11 &\footnotesize 121 &\footnotesize 0.0909
&\footnotesize 34713 &\footnotesize 243936 &\footnotesize 0.1423
&\footnotesize 0.1483  &\footnotesize 209  &\footnotesize 56\\
\footnotesize X03
&\footnotesize 12 &\footnotesize 121 &\footnotesize 0.0992
&\footnotesize 37280 &\footnotesize 243936 &\footnotesize 0.1528
&\footnotesize 0.0824  &\footnotesize 254  &\footnotesize 70\\
\footnotesize X04
&\footnotesize 11 &\footnotesize 121 &\footnotesize 0.0909
&\footnotesize 30519 &\footnotesize 243936 &\footnotesize 0.1251
&\footnotesize 0.0395  &\footnotesize 663  &\footnotesize 168\\
\footnotesize X05
&\footnotesize 10 &\footnotesize 121 &\footnotesize 0.0826
&\footnotesize 30878 &\footnotesize 243936 &\footnotesize 0.1266
&\footnotesize 0.1173  &\footnotesize 86   &\footnotesize 31\\
\footnotesize X06
&\footnotesize 10 &\footnotesize 121 &\footnotesize 0.0826
&\footnotesize 31133 &\footnotesize 243936 &\footnotesize 0.1276
&\footnotesize 0.0562  &\footnotesize 118  &\footnotesize 44\\
\footnotesize X07
&\footnotesize 13 &\footnotesize 121 &\footnotesize 0.1074
&\footnotesize 37463 &\footnotesize 243936 &\footnotesize 0.1536
&\footnotesize 0.0887  &\footnotesize 170  &\footnotesize 64\\
\footnotesize X08
&\footnotesize 12 &\footnotesize 121 &\footnotesize 0.0992
&\footnotesize 39287 &\footnotesize 243936 &\footnotesize 0.1611
&\footnotesize 0.0564  &\footnotesize 155  &\footnotesize 60\\
\footnotesize
&\footnotesize  &\footnotesize  &\footnotesize
&\footnotesize  &\footnotesize  &\footnotesize
&\footnotesize  &\footnotesize  &\footnotesize\\
\footnotesize Y01
&\footnotesize 31 &\footnotesize 483 &\footnotesize 0.0642
&\footnotesize 31505 &\footnotesize 324576 &\footnotesize 0.0971
&\footnotesize 0.1114  &\footnotesize 168  &\footnotesize 350\\
\footnotesize Y02
&\footnotesize 28 &\footnotesize 465 &\footnotesize 0.0602
&\footnotesize 28575 &\footnotesize 312480 &\footnotesize 0.0914
&\footnotesize 0.1330  &\footnotesize 171  &\footnotesize 390\\
\footnotesize Y03
&\footnotesize 30 &\footnotesize 465 &\footnotesize 0.0645
&\footnotesize 28651 &\footnotesize 312480 &\footnotesize 0.0917
&\footnotesize 0.1229  &\footnotesize 146  &\footnotesize 312\\
\footnotesize Y04
&\footnotesize 30 &\footnotesize 457 &\footnotesize 0.0656
&\footnotesize 29119 &\footnotesize 307104 &\footnotesize 0.0948
&\footnotesize 0.0752  &\footnotesize 219  &\footnotesize 489\\
\hline
\end{tabular*}
\end{table*}


All the traffic matrices in our experiments are decomposed into three sub-matrices.
According to the results in Table 3,
these sub-matrices indeed satisfy the hypotheses of the decomposition model (\ref{A+E+N}):

1. The ranks of the deterministic traffic matrices in the Abilene dataset are less than 13;
for the GEANT dataset, the ranks are less than 31.
The rank of each deterministic traffic matrix is less than $11\%$ of the rank of the corresponding original traffic matrix.
Therefore, all the decomposed deterministic traffic matrices are typical low-rank matrices;

2. In both datasets, the $l_{0}$-norm of one anomaly traffic matrix does not exceed 40000.
For each anomaly traffic matrix,
less than $17\%$ entries are non-zero entries.
Therefore, all the anomaly traffic matrices are typical sparse matrices;

3. For all the twelve traffic matrices used in our experiments,
the ratio of the Frobenius norm between the noise traffic matrix and the original traffic matrix is less than 0.18.
Hence, noise traffic matrices usually contribute only a small proportion of the total energy.

In addition, the number of iterations and the computational time needed in the implementation of the APG algorithm are quite acceptable.
Specifically, for all the traffic matrices in the Abilene dataset,
the computational time is less than three minutes;
the average computational time for the GEANT traffic matrices is longer,
but still less than nine minutes.

\newpage

\begin{figure}[!htbp]
\centering
\scalebox{0.7}[0.7]{\includegraphics*{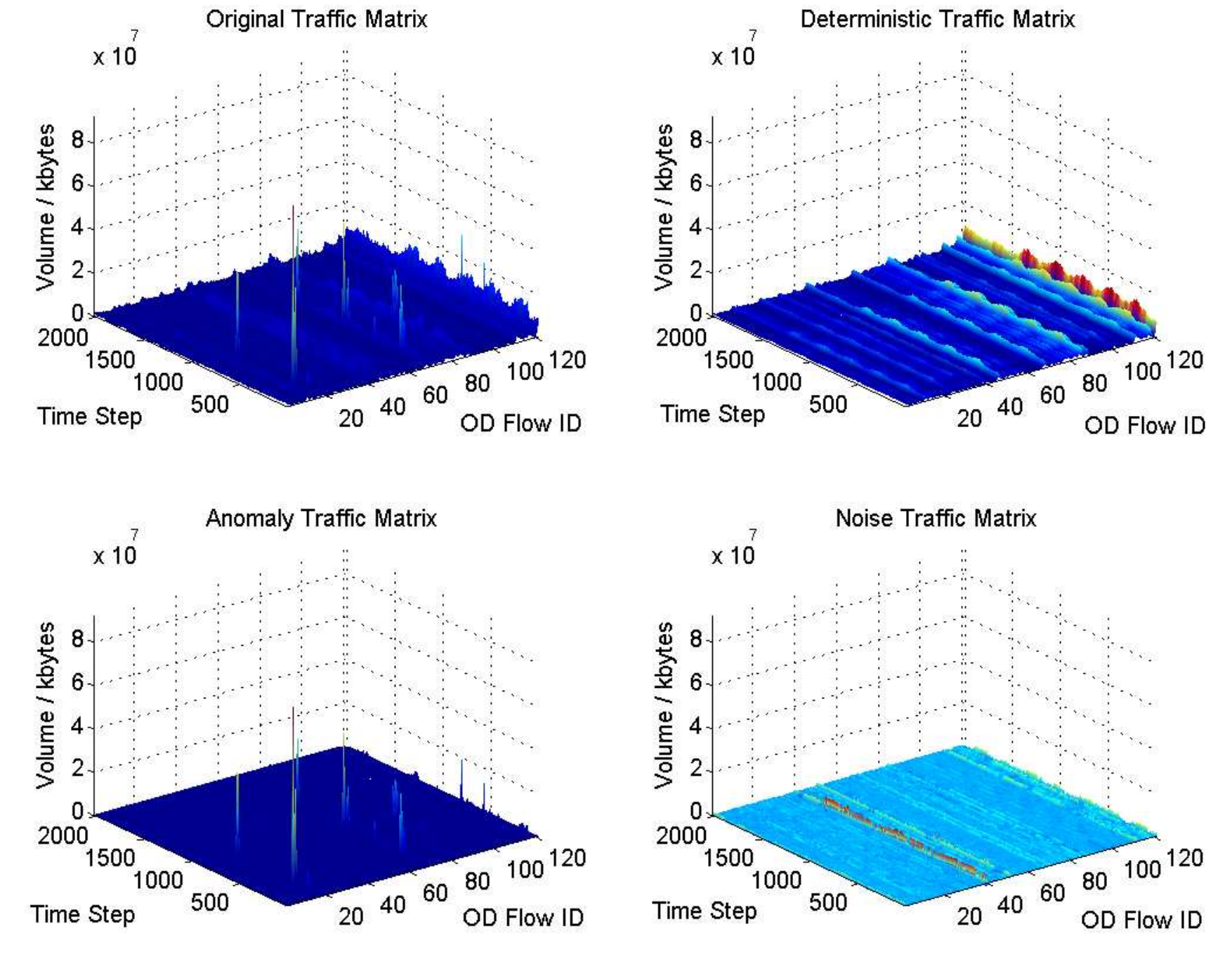}}
\begin{center}
\caption{The decomposition result of X01 by Relaxed PCP}\label{X01_PCP}
\end{center}
\medskip
\end{figure}

To further understand the experimental results,
we focus on the decomposition results of three traffic matrices:
$\textrm{X}01$, $\textrm{X}04$ and $\textrm{Y}01$.
Figure \ref{X01_PCP} shows the traffic matrix $\textrm{X}01$ (upper left) in the Abilene dataset,
and the three sub-matrices decomposed by the Relaxed PCP method.
We can see that the deterministic traffic matrix (upper right) contributes most of the energy of the total network traffic,
and the columns (corresponding to the deterministic traffic time series) show clear diurnal pattern,
especially for OD flows with large magnitude.
Most of the large volume anomalies in $\textrm{X}01$ are short-lived
and well isolated in the anomaly traffic matrix (bottom left).
This matrix is indeed quite sparse,
without distinct periodical traffic or noise traffic.
Furthermore, most of the entries in the noise traffic matrix (bottom right) have small magnitudes,
therefore the noise traffic indeed contributes little energy to the total network traffic.
For most of the OD flows,
the variances of the noise traffic are proportional to their mean volume, that is,
an OD flow with large magnitude usually has large noise traffic.
However, there also exist some OD flows of moderate magnitude which contain very large noise traffic,
which will be discussed in Section \ref{Noise Traffic Matrix}.
In summary, using the Relaxed PCP proposed described in section \ref{PCP_method},
we achieve a proper decomposition for the traffic matrix $\textrm{X}01$,
where all the sub-matrices satisfy the corresponding hypotheses in the decomposition model (\ref{A+E+N}).

\begin{figure}[!htbp]
\centering
\scalebox{0.7}[0.7]{\includegraphics*{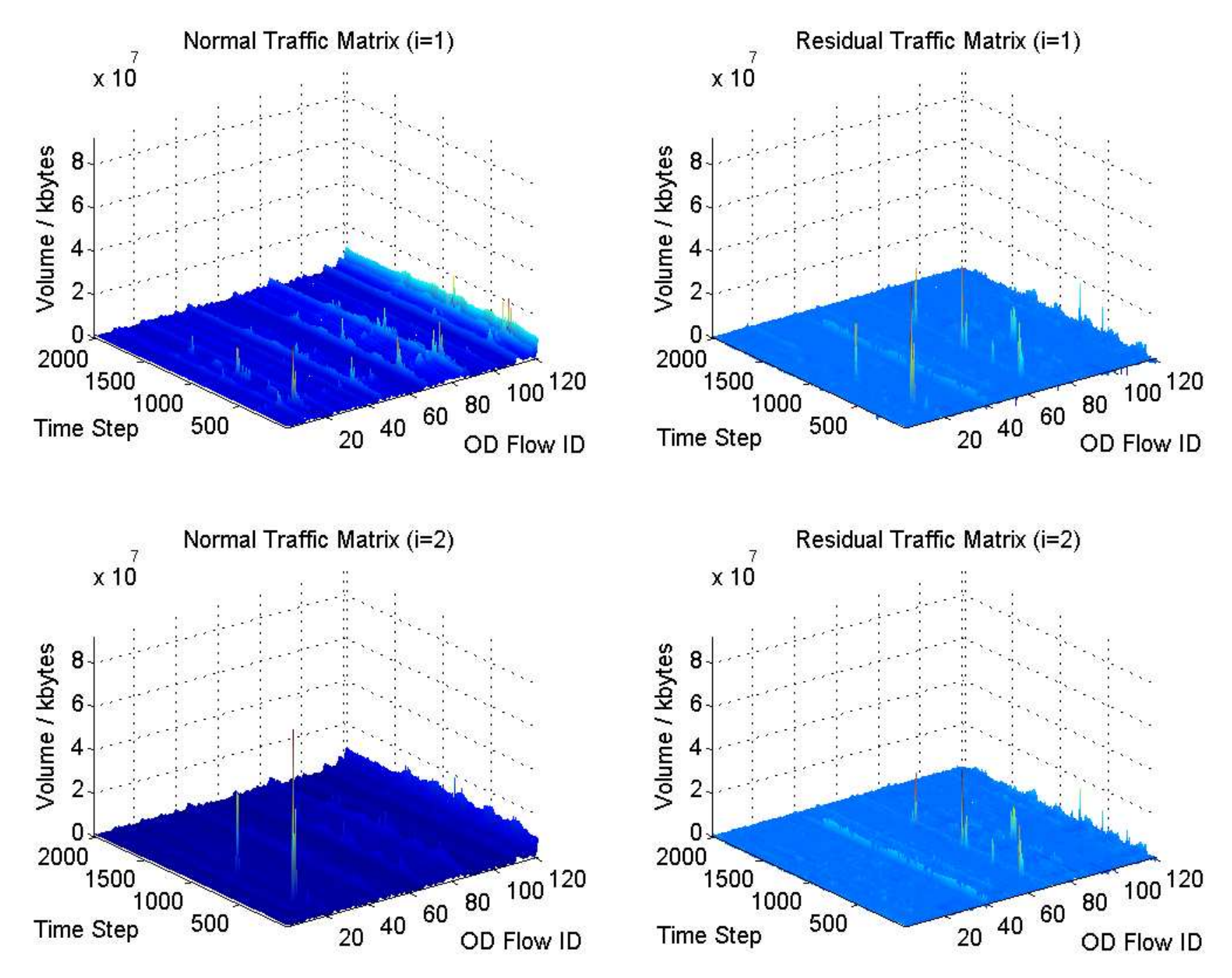}}\\
\scalebox{0.72}[0.72]{\includegraphics*{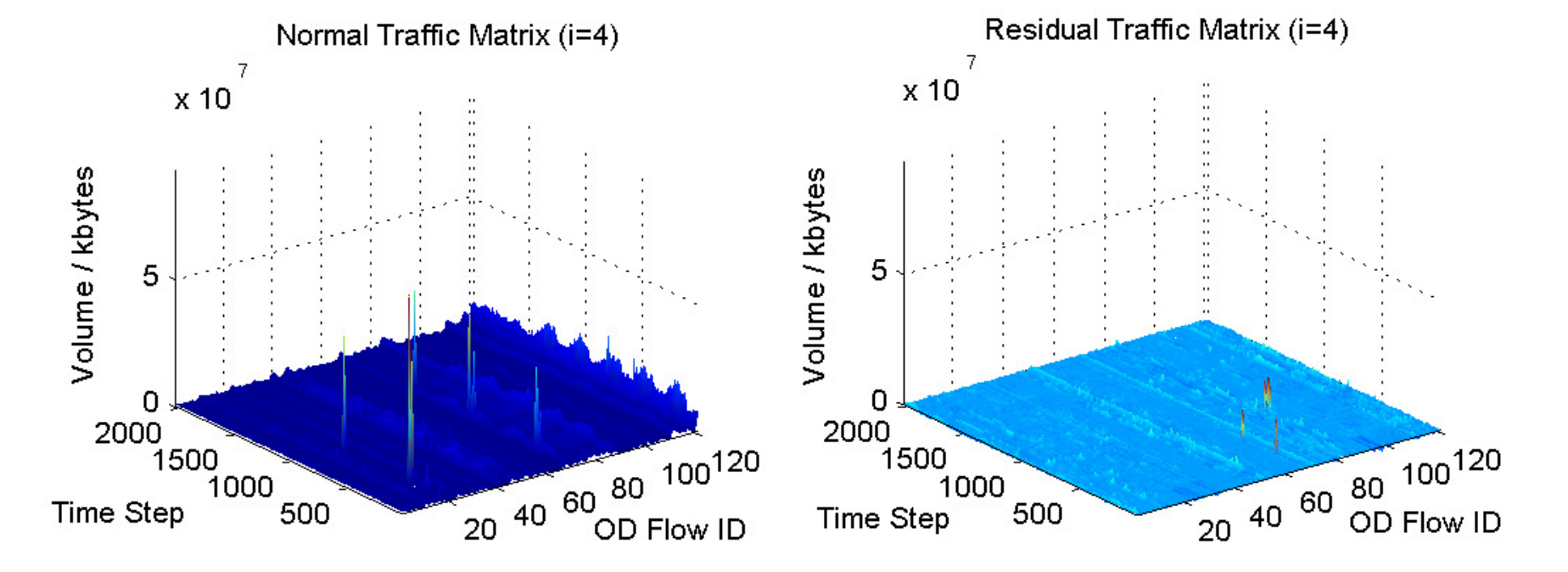}}
\begin{center}
\caption{The decomposition result of X01 by PCA}\label{X01_PCA}
\end{center}
\medskip
\end{figure}

For comparison,
we then decompose $\textrm{X}01$ using the PCA-based subspace method.
Recall that in Figure \ref{X01_eig} and Figure \ref{X01_eig_spe},
the first four eigenflows satisfy the criteria of both d-eigenflow and s-eigenflow,
and they contribute most of the energy of $\textrm{X}01$.
Therefore, the key is the classification results for these four eigenflows.
Suppose that for $i\in\{1,2,4\}$,
we classify the first $i$ eigenflow(s) as d-eigenflow,
and the rest (if exists) in the first four eigenflows as s-eigenflow.
The normal traffic matrix (projection onto the normal subspace) is then generated by the first $i$ d-eigenflow(s),
and the residual traffic matrix is the difference between the original traffic matrix and the normal traffic matrix.
In other words, the normal traffic matrix and the residual traffic matrix constitute a decomposition of $\textrm{X}01$.
For each choice of $i$,
the resulting normal traffic matrix and residual traffic matrix are shown in each row of Figure \ref{X01_PCA}.
We can see that for $i=1$,
most of the large anomaly traffic is isolated in the residual traffic matrix,
while the normal traffic matrix only captures partially the deterministic traffic in $\textrm{X}01$.
In other words, the residual traffic matrix also contains a large proportion of the diurnal traffic.
For $i=2$, although more diurnal traffic is present in the normal traffic matrix,
this matrix still contains large anomaly traffic.
Thus we can not efficiently identify large volume anomalies from the residual traffic matrix.
For $i=4$, since the fifth eigenflow is not a s-eigenflow and the sixth is one,
the normal traffic matrix is generated based on the first five eigenflows of $\textrm{X}01$.
In this case, the decomposition is even worse,
since most of the anomaly traffic is contained in the normal traffic matrix.
This again demonstrates that, for traffic matrices with large volume anomalies,
PCA is not a suitable method for the structural analysis problem.
Furthermore, we believe that its performance can not be significantly improved only by changing parameters or using heuristic mechanisms.

\begin{figure}[!htbp]
\centering
\scalebox{0.7}[0.7]{\includegraphics*{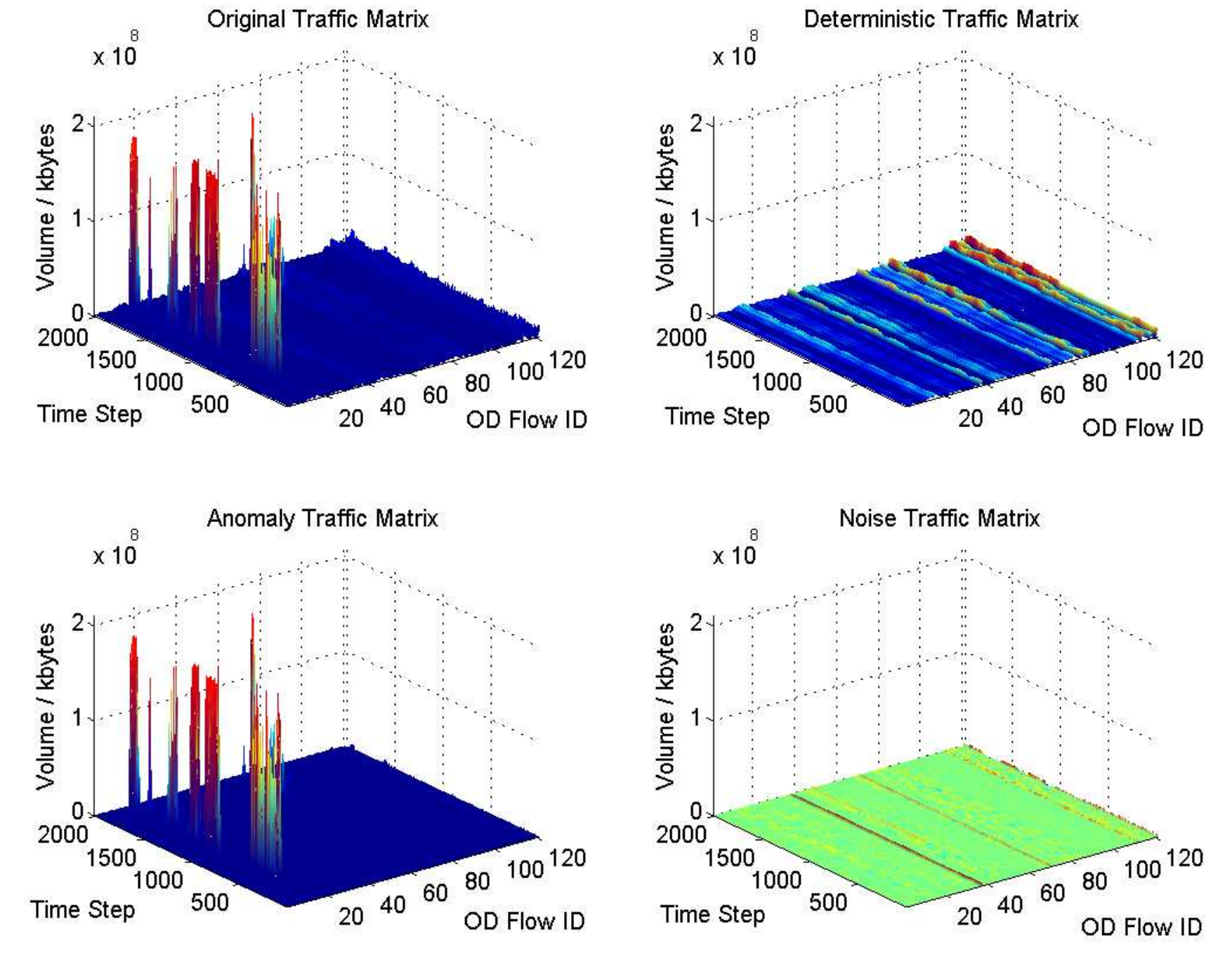}}
\begin{center}
\caption{The decomposition result of X04 by Relaxed PCP}\label{X04_PCP}
\end{center}
\medskip
\end{figure}

\begin{figure}[!htbp]
\centering
\scalebox{0.7}[0.7]{\includegraphics*{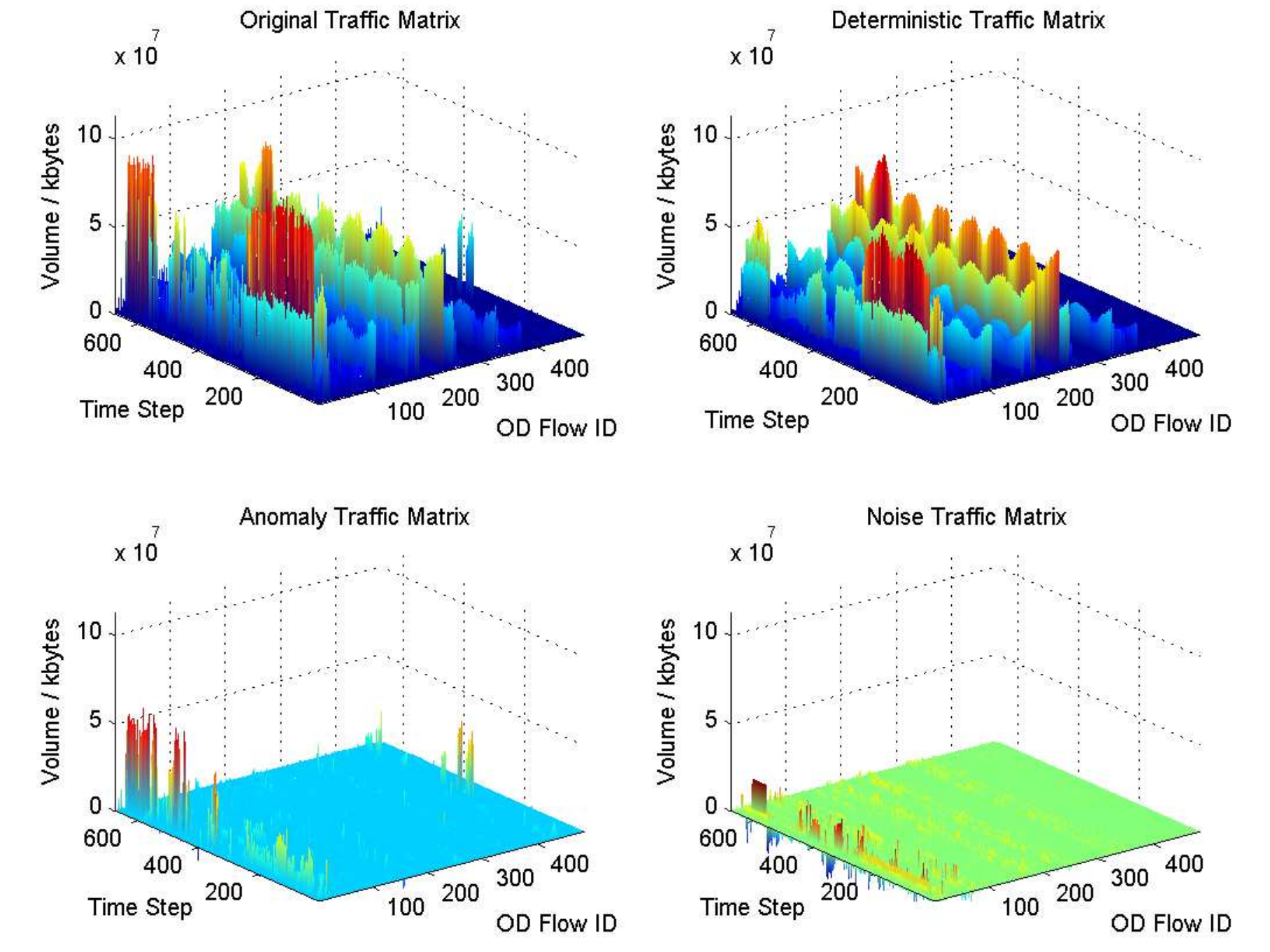}}
\begin{center}
\caption{The decomposition result of Y01 by Relaxed PCP}\label{Y01_PCP}
\end{center}
\medskip
\end{figure}

Figure \ref{X04_PCP} shows the traffic matrix $\textrm{X}04$ in the Abilene dataset,
as well as three sub-matrices decomposed by the Relaxed PCP method
(they are arranged in the same way as Figure \ref{X01_PCP}).
A significant difference between $\textrm{X}04$ and $\textrm{X}01$ is that the former one contains long-lived large volume anomalies.
Therefore, the anomaly traffic in $\textrm{X}04$ contributes a larger proportion of the total energy than that in $\textrm{X}01$.
Using the same ways to compute parameters $\lambda$ and $\mu$ in Algorithm \ref{algo1},
our experimental result in Figure \ref{X04_PCP} shows
that the Relaxed PCP method can also exactly decompose a traffic matrix with long-lived large volume anomalies.

Figure \ref{Y01_PCP} displays the decomposition result of the traffic matrix $\textrm{Y}01$ in the GEANT dataset.
In general, the result is similar to that of $\textrm{X}01$ and $\textrm{X}04$.
However, since the GEANT traffic matrices usually contain more unstable OD flows than the Abilene traffic matrices,
the periodical traffic pattern shown in the resulting deterministic traffic matrix
is less obvious compared to results of $\textrm{X}01$ and $\textrm{X}04$.

\newpage
\section{Discussions}\label{Discussions}
According to the traffic matrix decomposition model (\ref{A+E+N}),
we decompose the traffic matrix into three sub-matrices,
which correspond to three classes of network traffic.
Based on the experimental results obtained,
we now have further discussions on the deterministic traffic matrix and the noise traffic matrix in this section
(We do not discuss the anomaly traffic matrix since it may vary significantly for different input traffic matrices).

\subsection{Non-periodical Traffic in The Deterministic Traffic Matrix}\label{Deterministic Traffic Matrix}
As shown in Section \ref{Experiments},
for each traffic matrix in our experiments ($\textrm{X}01\sim\textrm{X}08$ and $\textrm{Y}01\sim\textrm{Y}04$),
the deterministic traffic matrix decomposed by algorithm \ref{algo1} has a low rank compared to the corresponding OD flow number.
In most cases, columns of the deterministic traffic matrix (deterministic traffic time series) display significant diurnal pattern.
However, there also exist several columns that contain traffic changes,
which are quite different from the periodical traffic.
This observation is quite obvious for the Abilene traffic matrices $\textrm{X}03$ and $\textrm{X}07$.

\begin{figure}[!htbp]
\centering
\scalebox{0.7}[0.7]{\includegraphics*{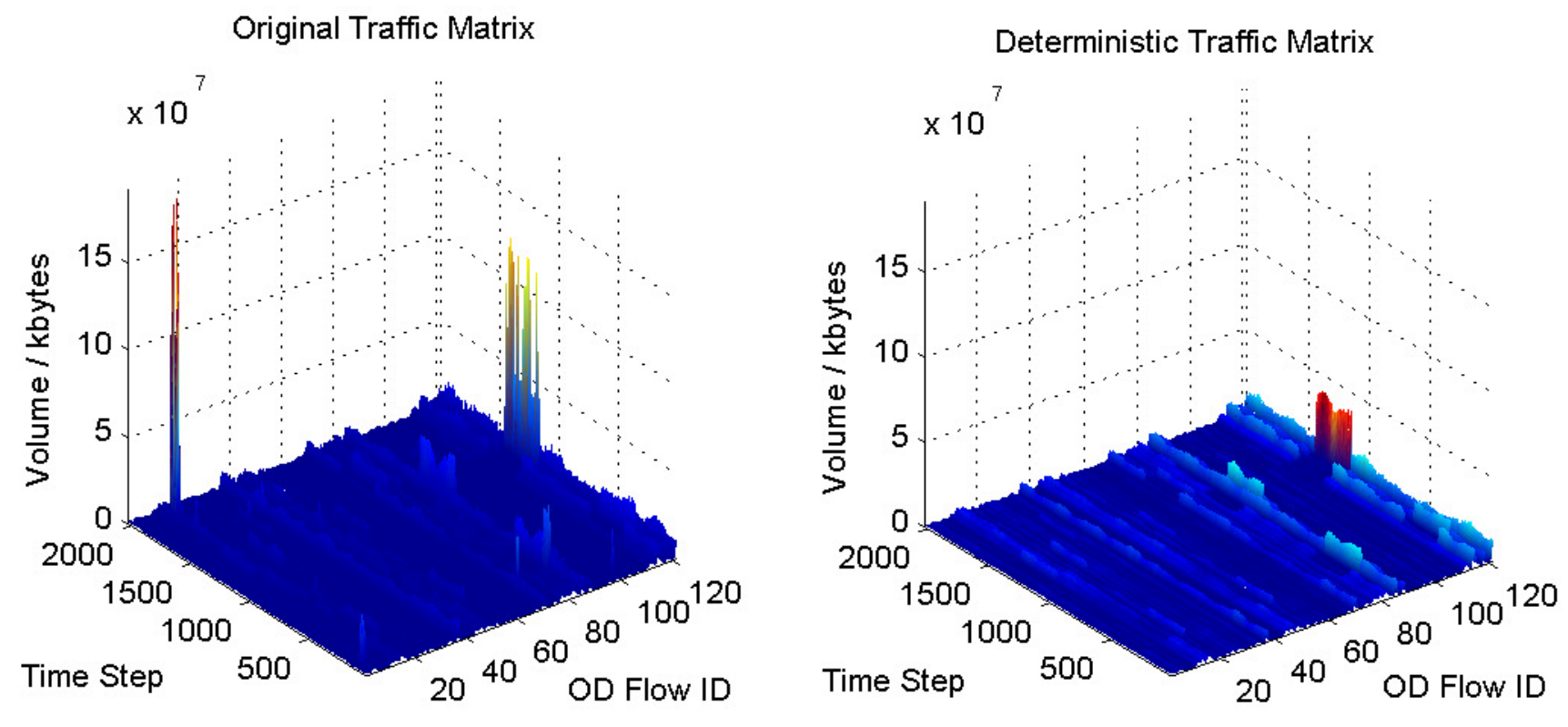}}
\begin{center}
\caption{The traffic matrix X03 (left) and the associated deterministic traffic matrix (right)}\label{X03_PCP}
\end{center}
\medskip
\end{figure}

\begin{figure}[!htbp]
\centering
\scalebox{0.75}[0.75]{\includegraphics*{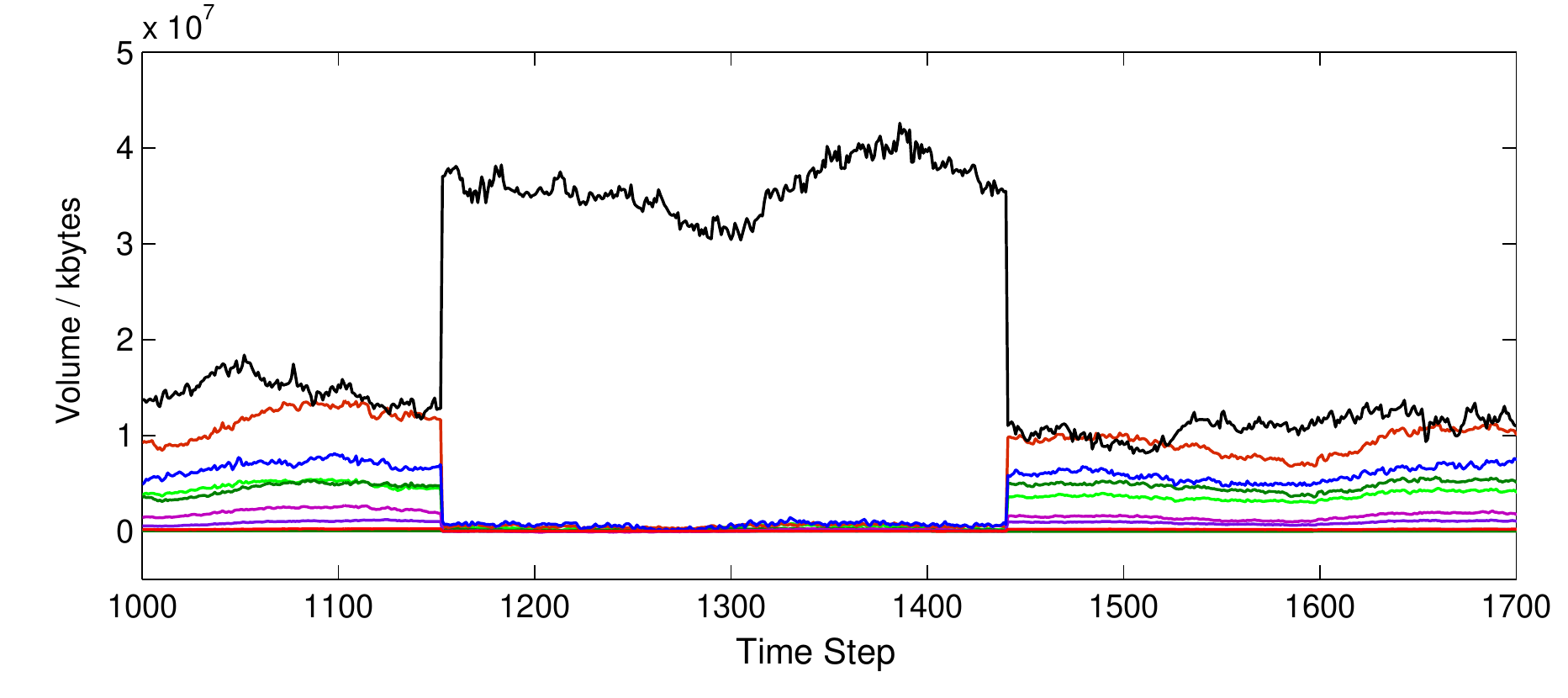}}
\begin{center}
\caption{Eleven columns of X03 (deterministic traffic time series) with the same source router "WASH"}\label{X03_ODflows}
\end{center}
\medskip
\end{figure}

As an example, Figure \ref{X03_PCP} displays $\textrm{X}03$ and the decomposed deterministic traffic matrix.
Clearly, we observe a few long-lived traffic changes in the deterministic traffic matrix.
These traffic changes affect a few columns (traffic time series) with long-lived growth or decline in terms of traffic volume,
and such growths and declines usually happen during the same time intervals.
In particular,
we illustrate in Figure \ref{X03_ODflows} eleven affected time series,
which have the same source router "WASH".
As we can see, one traffic growth and ten traffic declines all happen during the time intervals $[1150, 1450]$,
which share the same starting and ending time.
In fact, more than 20 time series in the deterministic traffic matrix are significantly affected,
but their source and destination routers do not present clear distribution laws.
These traffic changes have not been reported in the previous studies.
Therefore, it seems that the deterministic traffic matrix may contain non-periodical traffic changes,
which are usually combinations of long-lived traffic growths and declines during the same time intervals.
These changes can hardly be judged as any of the well known volume anomalies
such as DoS/DDoS, flash crowd, alpha, outages and ingress/egress shift \cite{Soule2}.
Since the Abilene traffic dataset only records OD flows' coarse-gained byte counts during every five-minute time interval,
and we do not have more detailed information about the network when these traffic changes happen,
it is difficult to explain the reason for these long-lived traffic changes. We leave this for future work.

\begin{figure}[!htbp]
\centering
\scalebox{0.75}[0.75]{\includegraphics*{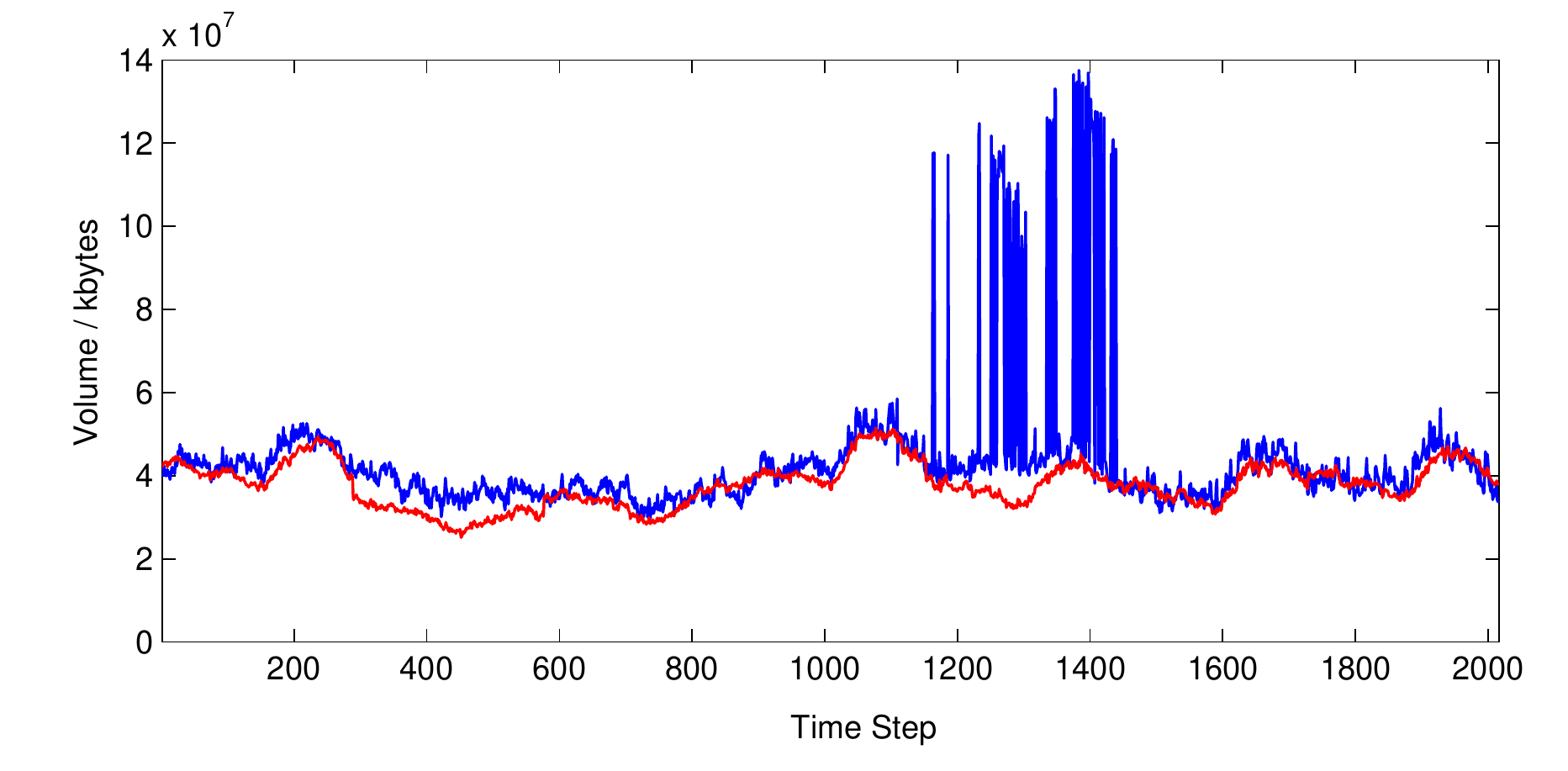}}
\begin{center}
\caption{The sum of eleven OD flow time series (blue) and the sum of the corresponding deterministic traffic series (red),
both with the same source router "WASH"}\label{X03_sum}
\end{center}
\medskip
\end{figure}

In addition,
we illustrate in Figure \ref{X03_sum} the sum of the aforementioned eleven OD flow time series in $\textrm{X}03$,
as well as the sum of the corresponding deterministic traffic series
(eleven columns of the deterministic traffic matrix of $\textrm{X}03$),
both with the same source router "WASH".
We can see that the sum of the OD flows contains some short-lived large traffic growths during the time intervals $[1150, 1450]$,
while these needle-like traffic growths can not be observed in the sum of the deterministic traffic series.
In fact, the latter sum presents typical periodical pattern during the whole week.
This shows that,
although individual deterministic traffic series with the same source router may contain significant traffic changes,
the sum of them tends to show expected patterns.
As a result, if we consider the total network traffic with the source router "WASH" (which is the sum of eleven OD flows),
the anomaly traffic component can be well decomposed by the Relaxed PCP method.

\subsection{Some Features of The Noise Traffic Matrix}\label{Noise Traffic Matrix}
\subsubsection{The Proportion of Noise Traffic in Different OD flows}\label{Noise Proportion}
\begin{figure}[!htbp]
\centering
\scalebox{0.75}[0.75]{\includegraphics*{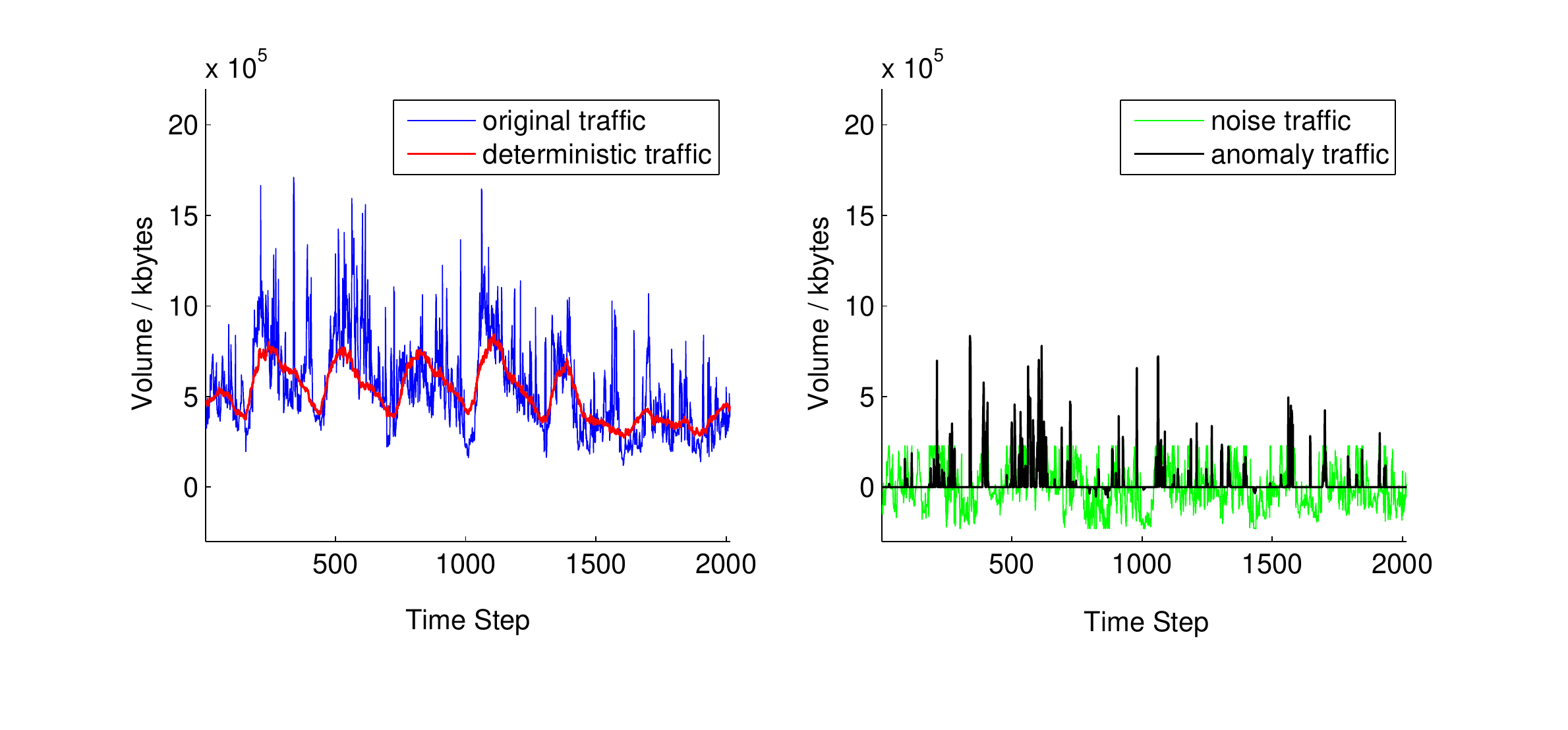}}
\begin{center}
\caption{The decomposition of OD flow time series No. 50 in X01.
Blue: original traffic; Red: deterministic traffic;
Black: anomaly traffic; Green: noise traffic }\label{X01_ODflow_deco1}
\end{center}
\medskip
\end{figure}

\begin{figure}[!htbp]
\centering
\scalebox{0.75}[0.75]{\includegraphics*{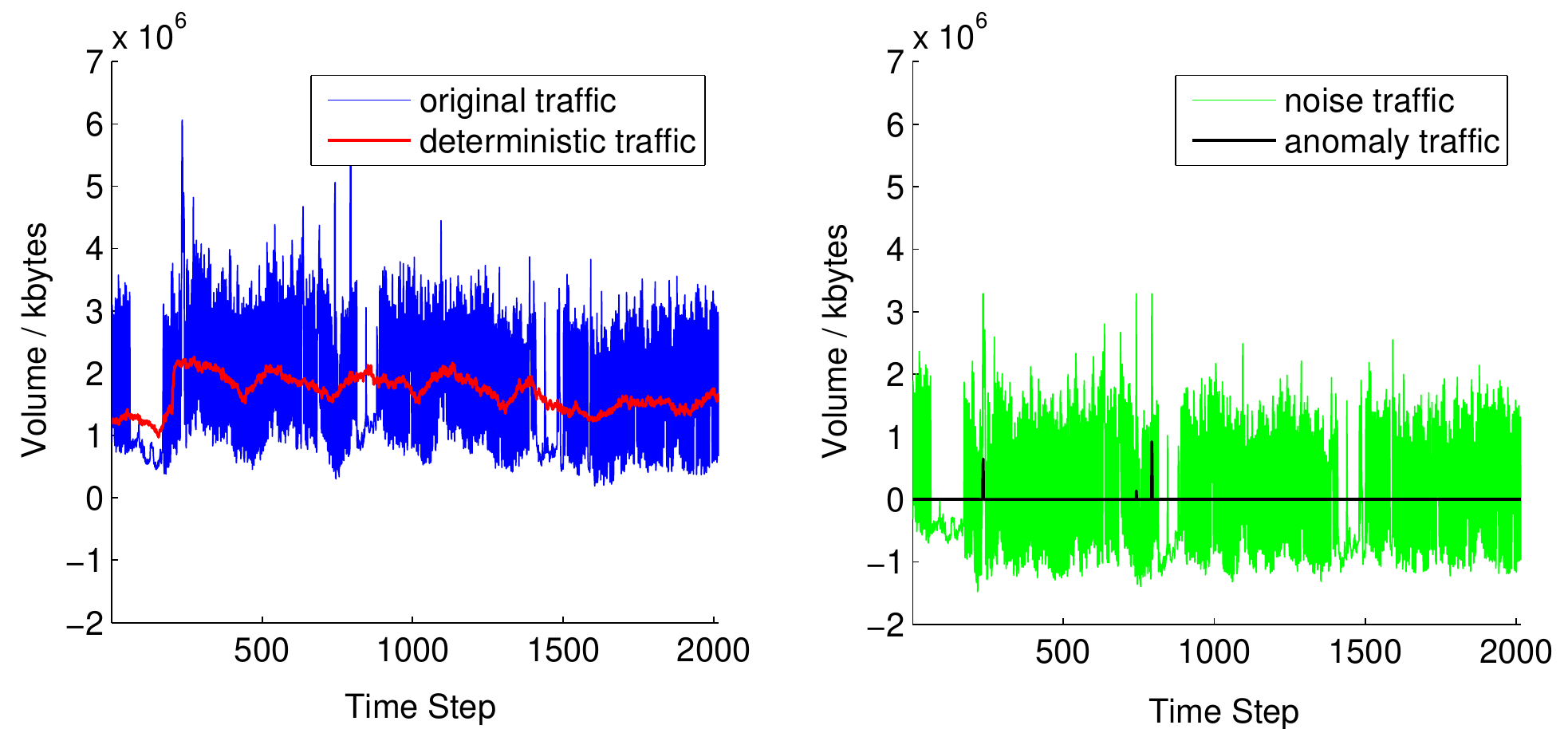}}
\begin{center}
\caption{The decomposition of OD flow time series No. 51 in X01.
Colors indicate the same classes of traffic as that in Figure \ref{X01_ODflow_deco1}}\label{X01_ODflow_deco2}
\end{center}
\medskip
\end{figure}

As can be seen from Table 3 in Section \ref{Experiments},
noise traffic matrices contribute a small proportion of the total network traffic.
However, we observe that the ratios of the noise traffic to the total traffic vary in different OD flows.
For instance, Figure \ref{X01_ODflow_deco1} and Figure \ref{X01_ODflow_deco2} illustrate decompositions of two OD flows in the Abilene traffic matrix $\textrm{X}01$, namely OD flow No. 50 and No. 51, respectively.
More specifically, for OD flow No. 50, the total traffic time series (blue) is mainly contributed
by the deterministic traffic time series (red) and the anomaly traffic time series (black),
and the noise traffic time series (green) has much smaller average magnitude.
Therefore, we conclude that the noise traffic is not an important component in OD flow No. 50.
In fact, this is the case for most of the OD flows.

However, a small number of OD flows have different decomposition results.
For example, for OD flow No. 51, the noise traffic time series has quite large average magnitude compared to the original traffic time series,
therefore the noise traffic becomes a significant component for this OD flow
hence should not be neglected in the analysis.
Actually, this OD flow contains large-amplitude oscillations,
which is not a common feature for all the OD flows,
and it should be classified as the noise traffic in the network.

To summarize, although noise traffic time series are usually small in magnitude,
they can not be neglected in the analysis of a few OD flows which contain large oscillations.

\subsubsection{The Variance of Noise Traffic Time Series}\label{Noise Variance}
\begin{figure}[!htbp]
\centering
\scalebox{0.8}[0.8]{\includegraphics*{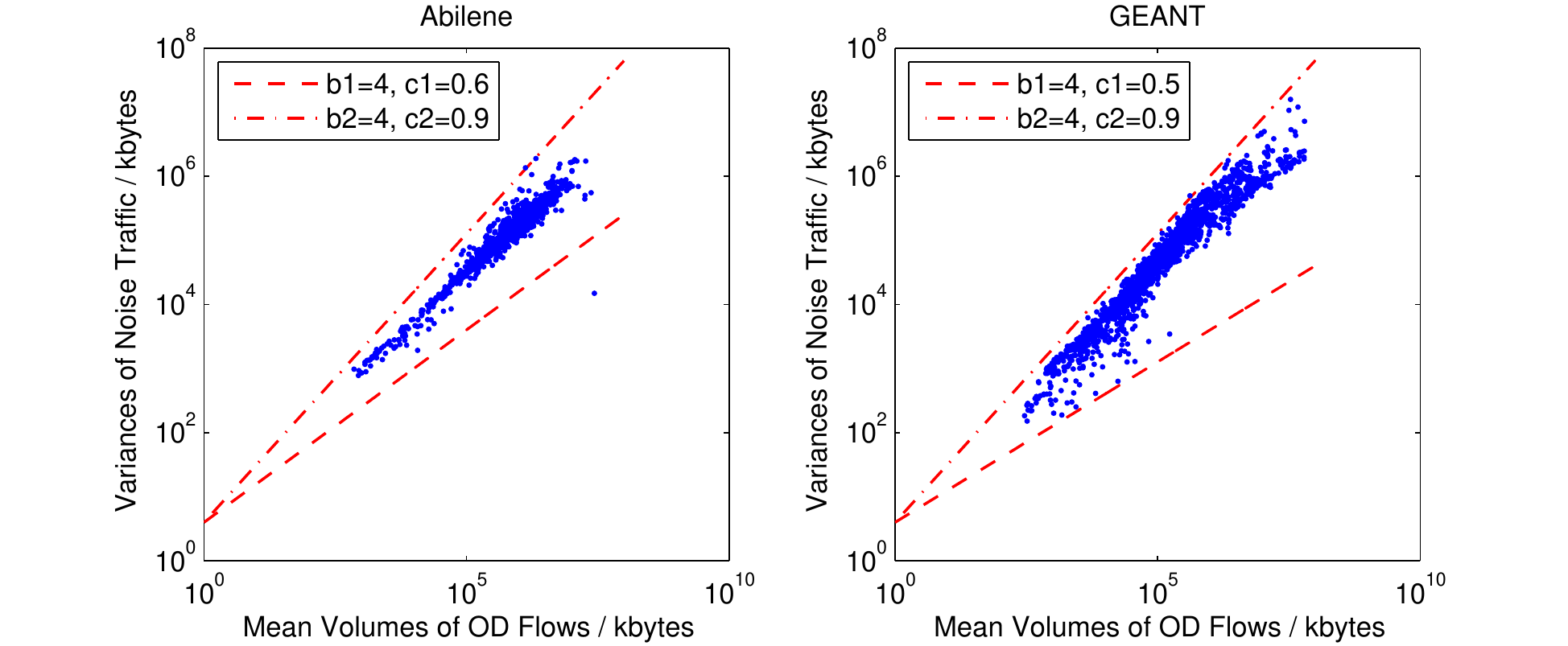}}
\begin{center}
\caption{The relationship between variance of the noise traffic and the mean volume of the corresponding OD flow.
Left: the Abilene networks; Right: the GEANT networks}\label{Gauss_var1}
\end{center}
\medskip
\end{figure}

The energy (variance) of the noise traffic may vary significantly in different OD flows.
Suppose that $X$ is a traffic matrix,
and we compose it as $X=A+E+N$ by Algorithm 1.
For each OD flow time series $X_{j}$ (one column of $X$),
we are interested in the relationship between variance $Var(N_{j})$ of the noise traffic
(which is estimated by the standard deviation of noise traffic time series $N_{j}$)
and statistics of $X_{j}$.
For example, Figure \ref{Gauss_var1} illustrates the relationship between $Var(N_{j})$ and $\mean(X_{j})$.
Specifically, for each data point in the figure,
the horizontal axis represents the mean value of an OD flow time series,
and the vertical axis represents the variance of the noise traffic of the same OD flow.
Here we have analyzed all the OD flows in our datasets,
where time series of the same OD flow in different weeks are considered as different data points.
Therefore, we have $121\times 8=968$ data points for the Abilene dataset,
and 1870 for the GEANT dataset.

It is clear from Figure \ref{Gauss_var1} that there is a strong positive correlation between the mean volume of OD flows and the variance of the corresponding noise traffic.
In the log-log plot,
the distribution of the data points follows a weak linear relationship,
and such relationship is more noticeable for the Abilene dataset.
Therefore,
it is reasonable to assume that in most cases
the variance of the noise traffic of an OD flow can be approximated by a power function of the mean volume of the OD flow,
which can be written as
\begin{equation}\label{power_function}
Var(N_{j})\approx b\mean(X_{j})^{c},
\end{equation}
where $b$ and $c$ are some positive parameters.
Notice that there exist many mathematical methods for the estimation of the parameters $b$ and $c$;
However, this is beyond the scope of the current study and we leave it for future work.
Instead, we propose empirical bounds for the variance of the noise traffic for the two datasets,
which are two parameter pairs $(b1, c1)$ and $(b2, c2)$ satisfying
\begin{equation}
b1\mean(X_{j})^{c1}\leq Var(N_{j})\leq b2\mean(X_{j})^{c2}.
\end{equation}
As labeled in Figure \ref{Gauss_var1},
for the Abilene dataset,
the choices $b1=b2=4$, $c1=0.6$ and $c2=0.9$ seem to work well for most of the data points except a few outliers;
for the GEANT dataset,
$b1=b2=4$, $c1=0.5$ and $c2=0.9$ are the reasonable choices.

In addition,
we have also analyzed the relationships between the variance of the noise traffic and several other statistics of the corresponding OD flow,
such as the $l_{2}$-norm, the median value, and the variance of the OD flow.
For all of them,
we have observed the positive correlation between the two as well,
but not as significant as the correlation between the noise variance and the mean volume of the flow.
In this case, it is less obvious to find an explicit mathematical model for the correlation as equation (\ref{power_function}).

\medskip

\begin{figure}[!htbp]
\centering
\scalebox{0.8}[0.8]{\includegraphics*{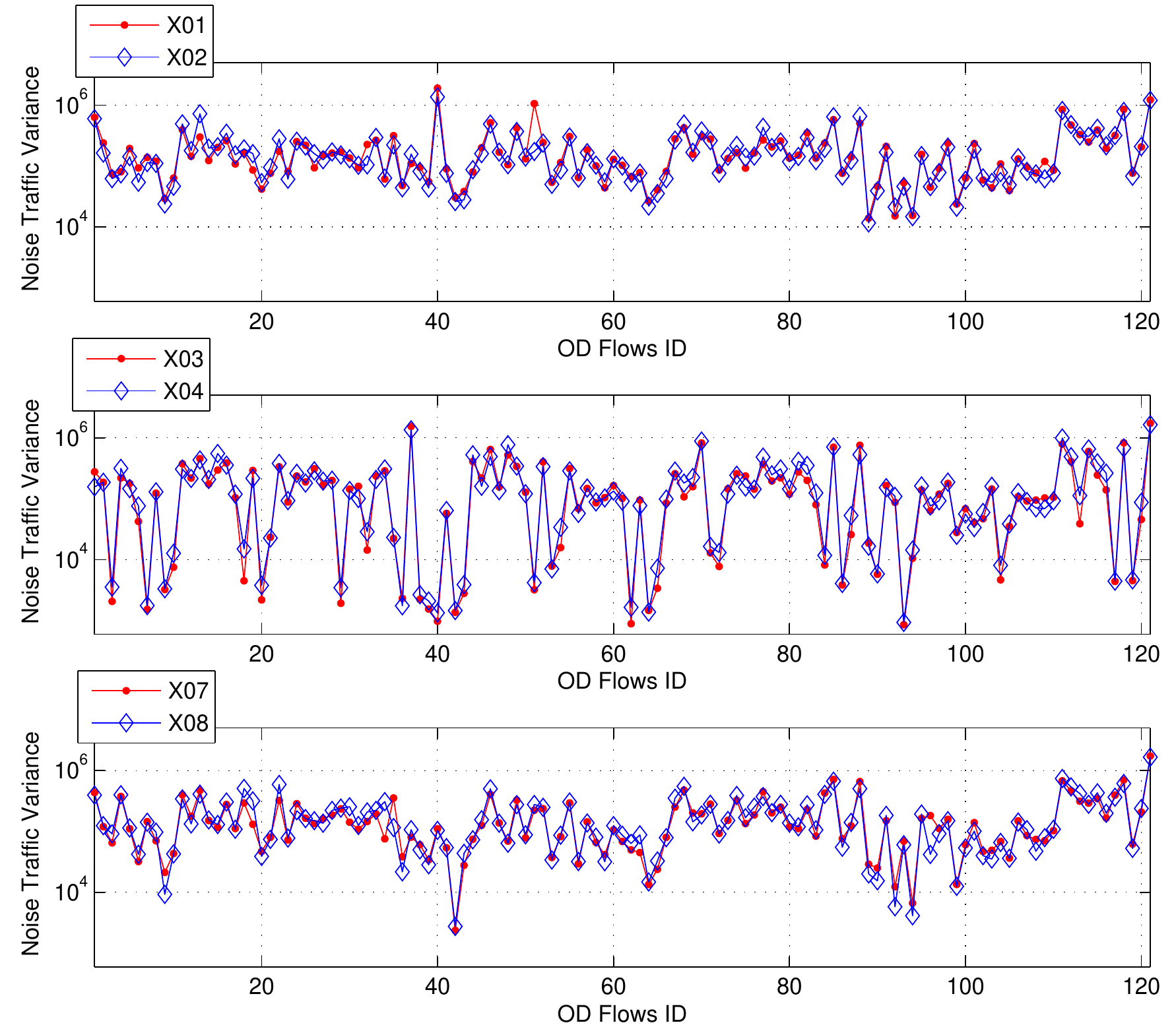}}
\begin{center}
\caption{The pair of variances of the noise traffic for two consecutive weeks}\label{temporal_stability}
\end{center}
\medskip
\end{figure}

Finally,
it is interesting to study the temporal stability of variances of the noise traffic during different weeks.
Suppose we have two traffic matrices that record the traffic volume of the same network for two consecutive weeks.
By decomposing them using the APG algorithm independently,
we first obtain two noise traffic matrices, one for each week.
Recall that each column vector of a noise traffic matrix represents the noise traffic time series of an OD flow.
For each OD flow,
we then compare the pair of variances of the corresponding noise traffic for the two consecutive weeks.
Specifically, we choose traffic matrices for three pairs of consecutive weeks in the Abilene dataset:
\begin{itemize}
  \item $\textrm{X}01$ (from March 1, 2004) and $\textrm{X}02$ (from March 8, 2004);
  \item $\textrm{X}03$ (from April 2, 2004) and $\textrm{X}04$ (from April 9, 2004);
  \item $\textrm{X}07$ (from May 8, 2004) and $\textrm{X}08$ (from May 15, 2004).
\end{itemize}
Since each Abilene traffic matrix contains 121 OD flows (columns),
the variances of the corresponding noise traffic include 121 data points.
Figure \ref{temporal_stability} shows the variances of the noise traffic for the six chosen traffic matrices
($121\times 6 = 726$ data points in total).

We observe the following features in Figure \ref{temporal_stability}:
(1) For each traffic matrix, the variances of the noise traffic of different OD flows may vary significantly;
(2) For traffic matrices of two consecutive weeks,
the variances of the noise traffic of the same OD flow are similar in most cases;
(3) For two traffic matrices that do not represent two consecutive weeks,
the variances of the noise traffic of the same OD flow may vary significantly
(Take the same OD flows in $\textrm{X}01$ and $\textrm{X}03$ as examples).

However, our observations are not sufficient to conclude that the variance of the noise traffic is strictly stable, the reason being that:
(1) The analysis above is not comprehensive enough as our datasets do not contain traffic matrices for many consecutive weeks;
(2) There also exist a few variance pairs in which one is obviously different from the other, although they correspond to the same OD flow for the two consecutive weeks.
We plan to investigate the unstable variances of the noise traffic in future work.

\section{Conclusions}\label{conclusions}
In this paper, we focus on the structural analysis of the traffic matrix that has been polluted by large volume anomalies.
We first demonstrate that the PCA-based analysis method performs poorly for polluted traffic matrices.
Next, we propose a new decomposition model for the traffic matrix that is more practical in the analysis of empirical network traffic data,
and study the decomposition problem using the Relaxed Principal Component Pursuit method.
Finally, we discuss the experimental results in more details for the deterministic and noise traffic matrix.
The major findings in this paper are:

1. We experiment the classical PCA method for traffic matrix analysis.
Different from the previous works \cite{Lakhina1}\cite{Lakhina2},
the traffic matrices that we analyze contain some large volume anomalies.
In this case, our results show that the eigenflow classification is neither complete nor orthogonal,
which suggests that PCA is unable to decompose accurately the traffic matrix into the normal traffic matrix and the large anomaly traffic matrix.

2. Based on the empirical network traffic data,
a new decomposition model for the traffic matrix is proposed in Section \ref{Composition Model}.
To the best of our knowledge, it is a novel way of formalizing the structure of the traffic matrix,
which also provides a simple view of the traffic matrix analysis problem.
Moreover, this model helps explain intuitively some of the limitations of the classical PCA method in our experiments.

3. According to the decomposition model of the traffic matrix,
we show that the problem of traffic matrix decomposition is equivalent to the robust PCA problem,
which has been extensively studied recently.
Based on the Relaxed Principal Component Pursuit method and the Accelerated Proximal Gradient algorithm,
we develop an algorithm for the decomposition of traffic matrices that may contain large volume anomalies.
The experimental results demonstrate the efficiency and flexibility of the proposed algorithm.

4. We discuss some detailed features of the deterministic traffic matrix and the noise traffic matrix.
Firstly, we observe that the deterministic traffic matrix may contain non-periodical traffic changes,
which are usually combinations of long-lived traffic growths and declines during the same time intervals.
Secondly, although the noise traffic matrix contributes a small proportion of the total network traffic in general,
the ratios of the noise traffic to the total traffic may vary significantly in different OD flows.
Thirdly, we find that there exists significant positive correlation between the mean volume of OD flow and the variance of the noise traffic time series,
and we further test the temporal stability of the variance of the noise traffic.

\medskip

To summarize, this paper is a preliminary study on applying the Relaxed PCP method for network traffic analysis,
whose efficiency and flexibility have been demonstrated in the experimental results.
For future work, we plan to further optimize the Relaxed PCP method to make it adaptable to the network traffic data,
and explore its applications in volume anomaly detection and data cleaning for the polluted traffic matrix.

\newpage
\section{Appendix}\label{Appendix}
In this appendix, we present the APG algorithm for traffic matrix decomposition.
\begin{algorithm}
\caption{\small APG for Traffic Matrix Decomposition}\label{algo1}
\textbf{Input}: the traffic matrix $X\in \mathbb{R}^{t\times p}$.

1. Compute the regularization parameter $\lambda$ using (\ref{lambda}).

2. Compute the Lagrangian parameter $\mu$ using (\ref{mu}) with $\sigma=1$.

3. For each OD flow time series $X_{j}$,
estimate the variance $\sigma_{j}$ of its Gaussian noise component using (\ref{sigma_j}).

4. Let $X=X/ diag\{\sigma_{j}\}$.

5. Let $A_{0}=A_{-1}=0$, $E_{0}=E_{-1}=0$, $t_{0}=t_{-1}=1$,

\quad $S_{1}^{A}=S_{1}^{E}=1$ and $k=0$.

6. \textbf{while} \ not converged \textbf{do}

\quad \quad $Y_{k}^{A}=A_{k}+\frac{t_{k-1}-1}{t_{k}}(A_{k}-A_{k-1})$;

\quad \quad $Y_{k}^{E}=E_{k}+\frac{t_{k-1}-1}{t_{k}}(E_{k}-E_{k-1})$;

\quad \quad $G_{k}^{A}=Y_{k}^{A}-\frac{1}{2}(Y_{k}^{A}+Y_{k}^{E}-X)$;

\quad \quad $G_{k}^{E}=Y_{k}^{E}-\frac{1}{2}(Y_{k}^{A}+Y_{k}^{E}-X)$;

\quad \quad $(U,S,V)=SVD(G_{k}^{A})$;

\quad \quad $A_{k+1}=U\mathcal{S}_{\frac{\mu}{2}}[S]V^{T}$;

\quad \quad $E_{k+1}=\mathcal{S}_{\frac{\lambda\mu}{2}}[G_{k}^{E}]$;

\quad \quad $t_{k+1}=\frac{1+\sqrt{4t_{k}^{2}+1}}{2}$;

\quad \quad $S_{k+1}^{A}=2(Y_{k}^{A}-A_{k})+(A_{k+1}+E_{k+1}-Y_{k}^{A}-Y_{k}^{E})$;

\quad \quad $S_{k+1}^{E}=2(Y_{k}^{E}-E_{k})+(A_{k+1}+E_{k+1}-Y_{k}^{A}-Y_{k}^{E})$;

\quad \quad $k=k+1$;

\quad \textbf{end while}

7. Let $X=X\cdot diag\{\sigma_{j}\}$.

\textbf{Output}:

$A=A_{k}\cdot diag\{\sigma_{j}\}$; $E=E_{k}\cdot diag\{\sigma_{j}\}$; $N=X-A-E$.
\end{algorithm}

\medskip

In Algorithm 1, $\mathcal{S}_{\varepsilon}[\cdot]: \mathbb{R}^{t\times p}\rightarrow \mathbb{R}^{t\times p}$ represents the soft-thresholding operator with parameter $\varepsilon>0$.
$\forall X\in\mathbb{R}^{t\times p}$,
$\mathcal{S}_{\varepsilon}[X]\in\mathbb{R}^{t\times p}$
and it satisfies
\begin{equation}
\mathcal{S}_{\varepsilon}[X](i,j)=
\begin{cases} X(i,j)-\varepsilon & \text{if $X(i,j)>\varepsilon$} \\ X(i,j)+\varepsilon & \text{if $X(i,j)<-\varepsilon$} \\ 0 & \text{otherwise}
\end{cases}.
\end{equation}
We choose the stopping criterion of Algorithm 1 as the one defined in \cite{Ma6},
which terminates the iterations when the quantity $\|S_{k+1}^{A}\|_{F}^{2}+\|S_{k+1}^{E}\|_{F}^{2}$ is less than a pre-defined tolerance parameter.

\section{Acknowledgment}\label{Acknowledgment}
We thank Professor Jinping Sun at Beihang University for his advices on the earlier draft of this paper.
We also thank Professor Pascal Frossard at Ecole Polytechnique F\'{e}d\'{e}rale de Lausanne (EPFL) for his help with the revised version of the paper.
Finally, we are grateful to the anonymous reviewers for their constructive suggestions on the paper.

\end{document}